\documentclass[aps,prapplied,reprint,amsmath,amssymb,superscriptaddress,longbibliography,floatfix]{revtex4-2}
\usepackage{graphicx,bm,capt-of}
\usepackage[T1]{fontenc}
\usepackage{lmodern}
\usepackage{hyperref}
\AtBeginDvi{}
\makeatletter
\let\CompanionOriginalFileLink\hyper@linkfile
\def\hyper@linkfile#1#2#3{%
  \begingroup\let\Hy@href@page\@empty
  \CompanionOriginalFileLink{#1}{#2}{#3}\endgroup}
\makeatother

\hypersetup{colorlinks=true,linkcolor=blue,citecolor=blue,urlcolor=blue}
\graphicspath{{figures/}}
\begin{document}
\title{Correlation-sensitive magic-state distillation from spin-squeezed neutral atoms}
\author{Ximo Wang}
\affiliation{College of Physics and Electronic Engineering, Shanxi University, Taiyuan, China}
\affiliation{Collaborative Innovation Center of Extreme Optics, Shanxi University, Taiyuan, China}
\author{Ruyue Guo}
\affiliation{College of Physics and Electronic Engineering, Shanxi University, Taiyuan, China}
\affiliation{Collaborative Innovation Center of Extreme Optics, Shanxi University, Taiyuan, China}

\author{Xinbo Li}
\affiliation{Collaborative Innovation Center of Extreme Optics, Shanxi University, Taiyuan, China}
\affiliation{Sanli Honors College, School of Physics and Electronics Engineering, Shanxi University, Taiyuan, People’s Republic of China}

\author{Min Feng}
\affiliation{College of Physics and Electronic Engineering, Shanxi University, Taiyuan, China}
\affiliation{Collaborative Innovation Center of Extreme Optics, Shanxi University, Taiyuan, China}
\author{Haozhe Wang}
\affiliation{College of Physics and Electronic Engineering, Shanxi University, Taiyuan, China}
\affiliation{Collaborative Innovation Center of Extreme Optics, Shanxi University, Taiyuan, China}
\author{Zhenqi Bai}
\affiliation{College of Physics and Electronic Engineering, Shanxi University, Taiyuan, China}
\affiliation{Collaborative Innovation Center of Extreme Optics, Shanxi University, Taiyuan, China}
\author{Yichi Zhang}
\email{zhangyichi@sxu.edu.cn}
\affiliation{College of Physics and Electronic Engineering, Shanxi University, Taiyuan, China}
\affiliation{Collaborative Innovation Center of Extreme Optics, Shanxi University, Taiyuan, China}
\begin{abstract}
We show that interblock correlations can reverse magic-state distillation even when every input exceeds the threshold for independent states. For the five-to-one Bravyi--Kitaev protocol with independent local Clifford twirls, we express the output in terms of measurable target-axis moments. Fourth-order correlations determine acceptance, while third- and fifth-order correlations determine the output polarization. This relation yields a direct sign test for distillation gain and opposite gains for inputs with identical single-qubit and two-qubit reduced states. We connect this mechanism to neutral-atom resource preparation through finite-range Ising dynamics. Block separation suppresses detrimental interblock coupling, and locally addressed Clifford phases release interaction-generated magic from a spin-squeezed precursor by changing the symmetry selected by code projection. A protocol that retains accepted source outputs achieves positive gain above a source-contrast threshold set by circuit errors and storage dephasing. The same moments specify a finite-sample experimental diagnostic. These results connect the correlation structure of a resource state to its preparation, acceptance, and distillation gain.
\end{abstract}
\maketitle
\raggedbottom

\section{Introduction}
Quantum error correction provides a route from coherent qubit control to reliable logical computation~\cite{shor1995scheme,steane1996error}. Magic-state distillation supplies the non-Clifford resources required for universality by converting noisy ancillas into states suitable for gate injection~\cite{bravyi2005universal}. Code constructions reduce the conversion overhead~\cite{bravyi2012magic,jones2013multilevel,meier2013magic}, while distillation limits and resource theories relate achievable output quality to nonstabilizer content and cost~\cite{campbell2010bound,veitch2014resource,howard2017application,wang2020efficiently,takagi2022one}. Factory designs organize these protocols into repeated resource production~\cite{litinski2019magic,gidney2019efficient,ogorman2017quantum}. Flag-based preparation~\cite{chamberland2019fault,chamberland2020very}, color-code structure~\cite{bombin2006topological,kubica2015unfolding}, transversal gates and gauge fixing~\cite{kubica2015universal,bombin2015gauge}, and code switching~\cite{anderson2014fault} provide complementary routes with distinct logical overheads~\cite{beverland2021cost}. Entangled inputs extend distillation to nonlocal magic, including resources generated by spin squeezing~\cite{bao2022magic}. Collective spin dynamics connects squeezing to entanglement~\cite{kitagawa1993squeezed,toth2009spin}; cavity-feedback theory and experiments~\cite{leroux2010implementation,schleiersmith2010squeezing}, together with Rydberg implementations in dipolar arrays, optical clocks, and dressed ensembles~\cite{bornet2023scalable,eckner2023realizing,hines2023spin}, establish physical methods for preparing such states. Experiments have progressed from NMR distillation~\cite{souza2011experimental} and fault-tolerant ion-trap preparation and injection~\cite{postler2022demonstration} to encoded superconducting resources~\cite{ye2023logical,gupta2024encoding}, error-suppressed surface-code memories~\cite{google2025below}, logical distillation in neutral atoms~\cite{rodriguez2025experimental}, and cultivation followed by code switching into a surface code~\cite{rosenfeld2026cultivation}. For neutral atoms, assembly and programmable geometry~\cite{barredo2016atom,endres2016atom,ebadi2021quantum}, coherent transport~\cite{bluvstein2022quantum,bluvstein2023logical}, Raman driving~\cite{levine2022dispersive}, parallel entangling gates~\cite{levine2019parallel,evered2023high}, and time-optimal gate designs~\cite{jandura2022time} combine with midcircuit cavity measurements and protected readout~\cite{deist2022mid,lis2023midcircuit,norcia2023midcircuit}. These advances support neutral-atom architectures that combine resource preparation with logical processing~\cite{bluvstein2025fault}.
However, parallel preparation exposes resource blocks to shared interactions and control environments. Finite-range Rydberg-dressed interactions~\cite{hines2023spin} can connect neighboring blocks when their separation is comparable to the interaction range. Scattered photons from midcircuit imaging disturb nearby data atoms~\cite{graham2023midcircuit}, and a shared magnetic field produces the correlated phase errors detected and corrected with spectator qubits~\cite{singh2023mid}. The resulting joint error statistics enter distillation through postselection. We find that the output fidelity can fall below the mean input fidelity even when every input exceeds the independent-state threshold. In finite-range atom-array calculations, replacing the correlated register by the product of its own reduced states restores positive gain while preserving each local input. This comparison identifies higher-order interblock correlations as the source of the reversal.

In this paper, we develop two methods that connect this effect to spin-squeezed resource preparation. Geometric separation suppresses coupling between blocks while retaining the intrablock interaction used to generate the resource. Locally addressed Clifford phase gates change the parity selected by code projection, allowing an Ising-prepared spin-squeezed state to supply encoded magic. We derive the correlated-input Bravyi--Kitaev map after independent local Clifford twirling and obtain a measurable criterion for the sign of the distillation gain. Inputs with identical single-qubit and two-qubit reduced states can have opposite gains because their higher-order moments differ. Finite-range simulations then relate the criterion to block geometry and phase control. A circuit model that retains accepted source outputs incorporates readout errors, loss, and storage histories and identifies the source-contrast threshold for positive gain. The resulting measurement protocol tests the gain directly from target-axis moments. Interblock separation and local phase control thus retain the useful intrablock interaction and reduce the correlations that degrade joint distillation.

\begin{figure*}[!t]
\centering
\includegraphics[width=\textwidth]{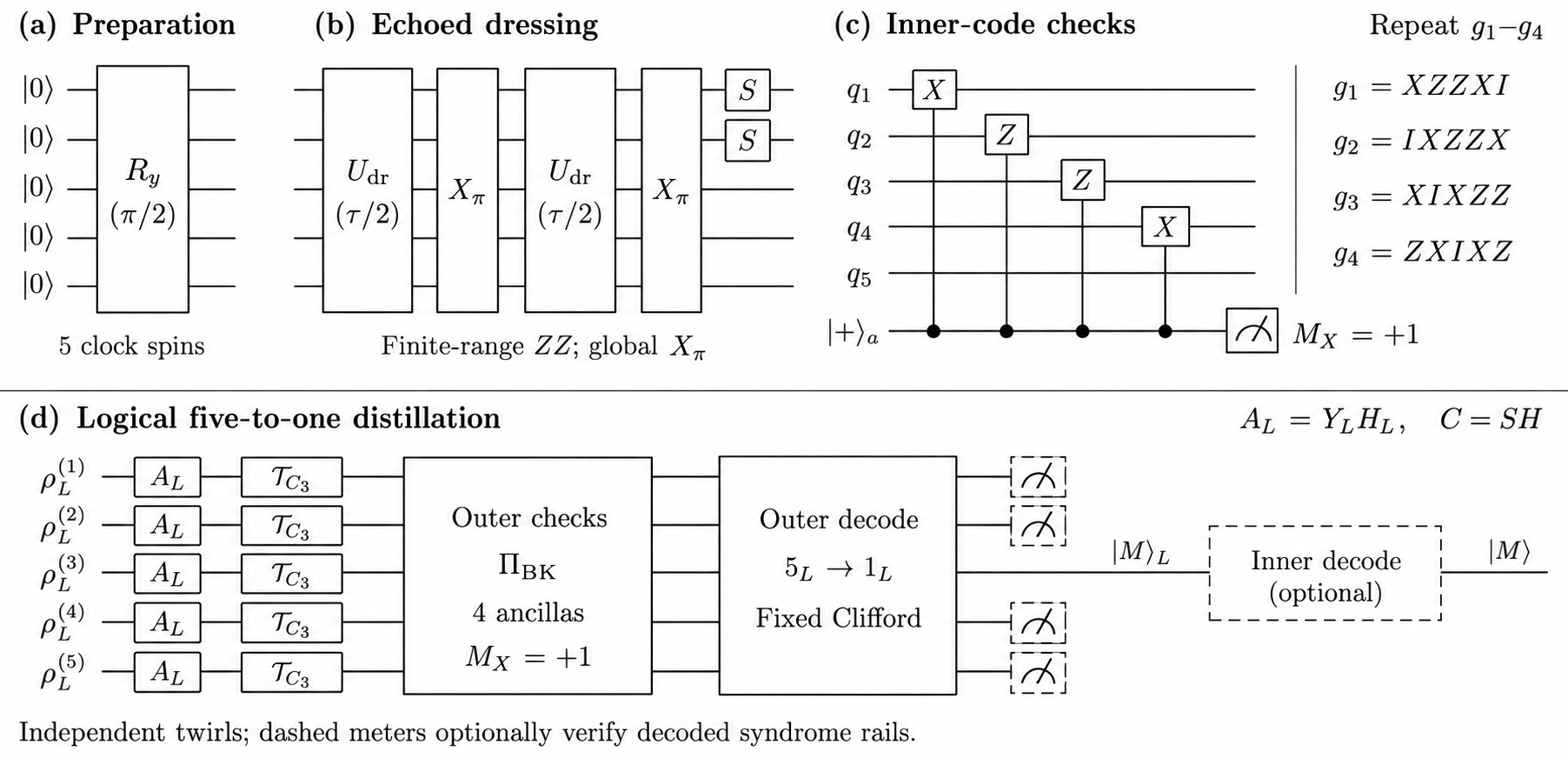}
\caption{Preparation and five-to-one magic-state distillation. (a) A collective $R_y(\pi/2)$ pulse prepares five clock-state atoms, initially in $|0\rangle^{\otimes5}$, in $|+\rangle^{\otimes5}$. (b) Two half-duration dressing intervals and two global $X_\pi$ pulses generate finite-range Ising interactions while refocusing static single-particle $Z$ shifts. Addressed $S$ gates are then applied to sites 1 and 2. (c) Ancilla-assisted extraction of an inner-code stabilizer, illustrated for $g_1=XZZXI$. An ancilla initialized in $|+\rangle$ controls the indicated Pauli factors and is measured in the $X$ basis. Repeating the check for $g_1,\ldots,g_4$ and retaining all four $+1$ outcomes projects the atoms into the $[[5,1,3]]$ code space. (d) Five accepted logical inputs, which may remain mutually correlated, undergo alignment $A_L=Y_LH_L$ and independently sampled cyclic Clifford twirls $\{I,C_L,C_L^2\}$, with $C_L=S_LH_L$. These twirls preserve joint target-axis correlations. Outer BK projection retains the all-plus syndrome; outer decoding and the fixed correction $XS$ then produce one encoded resource. The four outer ancilla preparations and readouts occur within the check module; a reset ancilla may be reused. Dashed meters indicate optional verification of the decoded syndrome rails. Optional inner decoding returns a physical qubit. The labels $|M\rangle_L$ and $|M\rangle$ denote the encoded and physical BK-type target direction $|T_{\rm BK}\rangle$. Operator products act from right to left: $A_L$ applies $H_L$ then $Y_L$, and $XS$ applies $S$ then $X$. Section~S3 of the Supplemental Materials specifies the physical gate schedules and Clifford frames.}
\label{fig:circuit}
\end{figure*}

\begin{figure*}[!t]
\centering
\includegraphics[width=\textwidth]{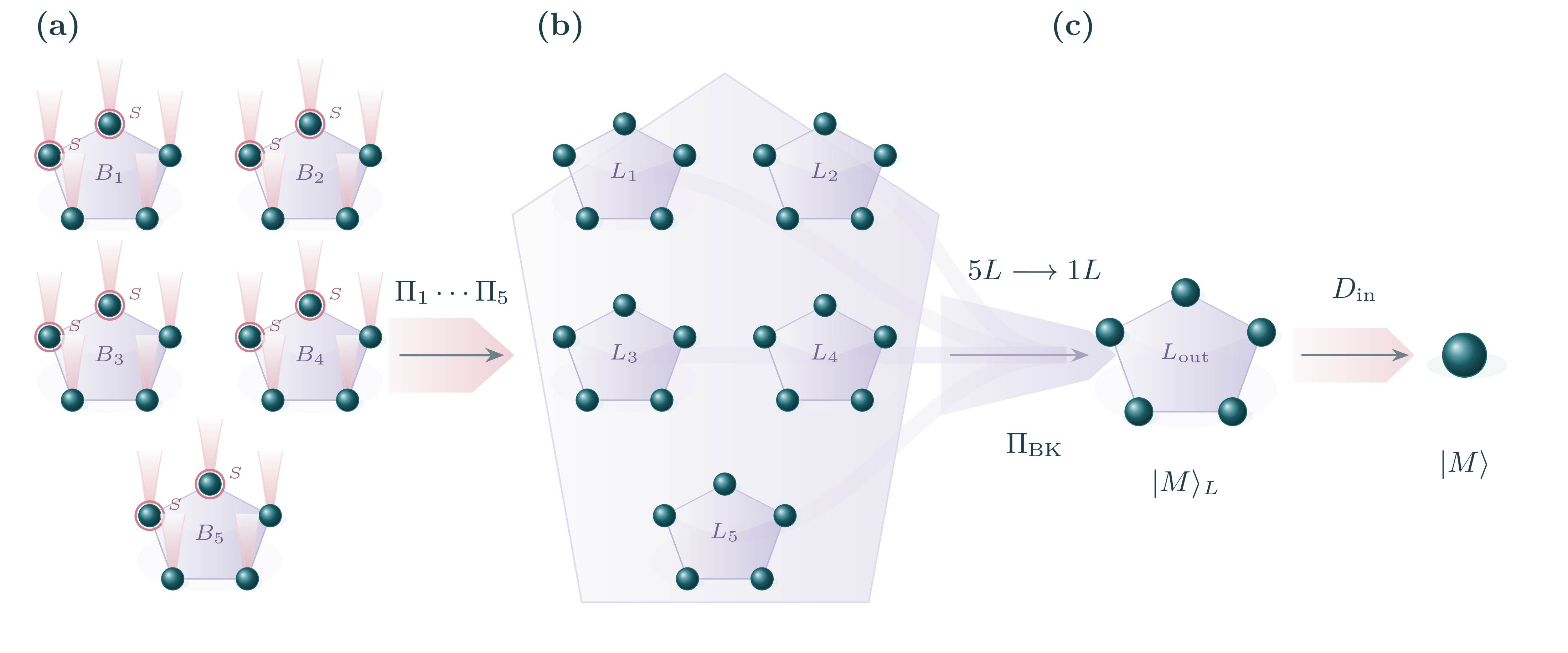}
\caption{Encoding hierarchy. (a) Five blocks $B_b$, each containing five physical atoms, receive Ising evolution and two addressed $S$ gates. (b) Joint inner acceptance $\Pi_1\cdots\Pi_5$ produces five encoded inputs $L_b$. (c) Alignment, independent twirls, outer BK projection, decoding, and the fixed Clifford correction produce one encoded resource; optional inner decoding $D_{\rm in}$ returns a physical qubit. Pentagon sheets indicate code membership, and state symbols indicate the target direction. The simulated positions form a one-dimensional chain. Single-shot joint acceptance is $p_{\rm tot}=p_{\rm in}^{(5)}p_{\rm BK}$, with the inner probability evaluated on the joint source.}
\label{fig:hierarchical-distillation}
\end{figure*}

\section{Distillation beyond the independent-input threshold}
\label{sec:correlation}
Figures~\ref{fig:circuit} and \ref{fig:hierarchical-distillation} summarize the preparation and encoding architecture. Each five-atom patch is projected into the $[[5,1,3]]$ code~\cite{laflamme1996perfect}; five accepted logical qubits then enter the outer BK protocol. We use the BK $T$-type state $|T_{\rm BK}\rangle$, with Bloch vector $(1,1,1)/\sqrt3$, as the target. Throughout, fidelity means the squared pure-state overlap $F=\langle T_{\rm BK}|\rho|T_{\rm BK}\rangle$, and distillation gain means output fidelity minus mean input fidelity. The source model, Clifford circuits, and conversion to gate-injection resources are developed in the Supplemental Materials (SM).

\subsection{The exact correlated-input map}
The $[[5,1,3]]$ code is generated by $g_1=XZZXI$, $g_2=IXZZX$, $g_3=XIXZZ$, and $g_4=ZXIXZ$. Its code-space projector and logical Pauli operators are
\begin{equation}
 \begin{aligned}
 \Pi&=\prod_{a=1}^{4}\frac{I+g_a}{2}
     =\frac1{16}\sum_{G\in\mathcal S}G,\\
 (\overline X,\overline Y,\overline Z)
 &=(X^{\otimes5},Y^{\otimes5},Z^{\otimes5}).
 \end{aligned}
 \label{eq:five-projector}
\end{equation}
where $\mathcal S$ is the stabilizer group. For an arbitrary joint input $\rho$, a Clifford decoder $V$ gives the unnormalized accepted output $\mathcal D(\rho)=\operatorname{Tr}_{1\ldots4}(V\Pi\rho\Pi V^\dagger)$. The success probability and raw logical Bloch components follow directly from stabilizer expectation values:
\begin{equation}
 p_{\rm acc}=\operatorname{Tr}(\rho\Pi),\qquad
 r_a^{\rm raw}=\frac{\operatorname{Tr}(\rho\Pi\overline\sigma_a)}
 {p_{\rm acc}}.
 \label{eq:general-bloch}
\end{equation}
These expressions apply to product states, classical mixtures, and entangled inputs. Conditioning on success introduces the normalization by $p_{\rm acc}$.

Let $M=(X+Y+Z)/\sqrt3$ and $\rho_s=(I+sM)/2$ for $s=\pm1$. Independently applying a uniformly random cyclic Clifford to each input dephases that input in the eigenbasis of $M$. The joint state becomes
\begin{equation}
 \rho_{\rm tw}=\sum_{\boldsymbol s\in\{\pm1\}^5}
 P(\boldsymbol s)\bigotimes_{i=1}^{5}\rho_{s_i},\qquad
 A_k=\sum_{|S|=k}\left\langle\prod_{i\in S}s_i\right\rangle.
 \label{eq:sign-mixture}
\end{equation}
Here $P(\boldsymbol s)$ is the joint probability distribution of the target-axis signs. It preserves their correlations. The quantity $A_k$ is the unnormalized sum of all $k$-qubit correlators along the target axis.

The stabilizer projector contains weight-four Pauli strings. Its logical Pauli cosets contain weight-three and weight-five strings. Their expectations select the corresponding correlation orders in Eq.~(\ref{eq:sign-mixture}). After the fixed output Clifford $XS$, the exact distillation map is
\begin{equation}
 \boxed{\begin{aligned}
 p_{\rm acc}&=\frac{3+A_4}{48},\\
 t_{\rm out}&=\frac{A_3-2A_5}{3+A_4}
 \end{aligned}}\qquad
 F_{\rm out}=\frac{1+t_{\rm out}}2.
 \label{eq:moment}
\end{equation}
where $t_{\rm out}=\langle M\rangle_{\rm out}$. The stabilizer expansion is given in SM Sec.~S1. Fourth-order correlations control acceptance, while the competition between third- and fifth-order correlations controls the accepted polarization. This identifies the information needed to predict distillation directly from the joint input state.

\subsection{Why individual inputs can pass and the joint input fail}
For identical independent inputs with $t=\langle s_i\rangle$, the correlation sums factor as $A_k=\binom5k t^k$. Equation~(\ref{eq:moment}) reduces to
\begin{align}
 p_{\rm acc}^{\rm iid}&=\frac{3+5t^4}{48},\qquad
 t_{\rm out}^{\rm iid}=\frac{2t^3(5-t^2)}{3+5t^4},\nonumber\\
 t_{\rm out}^{\rm iid}-t
 &=\frac{t(1-t^2)(7t^2-3)}{3+5t^4}.
 \label{eq:iid-map}
\end{align}
The independent-input fidelity threshold is therefore $F_{\rm th}=(1+\sqrt{3/7})/2$. For correlated inputs with the same local polarization, the exact gain condition becomes $A_3-2A_5>t(3+A_4)$. For unequal inputs, replacing $t$ in this comparison by their mean polarization tests gain relative to the mean input fidelity.

\begin{figure*}[!t]
\centering
\includegraphics[width=\textwidth]{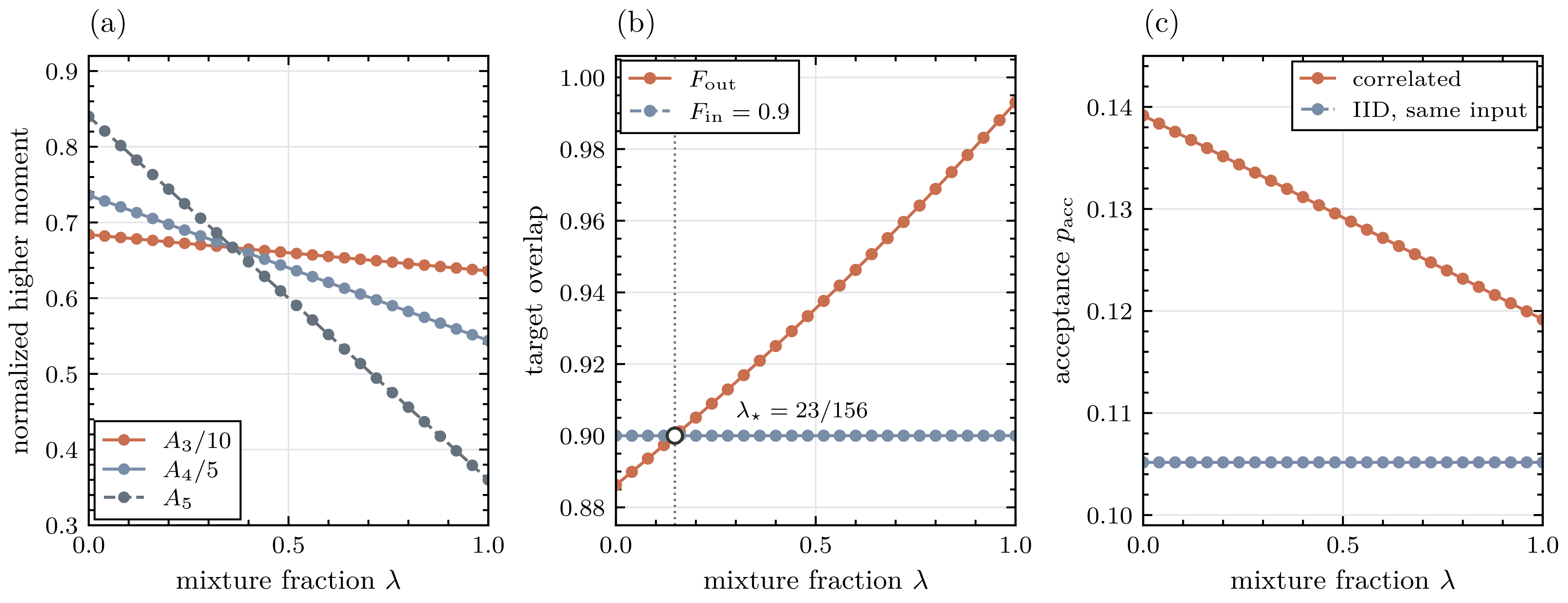}
\caption{Changing distillation gain at fixed local and pair data. The exchangeable family in Eq.~(S5) has fixed single-qubit and two-qubit reduced states. (a) Correlation sums normalized as $A_k/\binom5k$. (b) Accepted target fidelity, crossing the common input as the higher moments vary. (c) Success probability, compared with independent and identically distributed (IID) inputs of the same fidelity. Every mixture is evaluated by direct projection of all target-axis sign patterns.}
\label{fig:moment-sensitivity}
\end{figure*}

Figure~\ref{fig:moment-sensitivity} isolates the consequences of this condition. The entire family has the same single-qubit states and the same two-qubit reduced states. Varying only the higher-order error statistics moves the accepted output from below the common input fidelity to above it, and ultimately above the independent-input reference. Thus the same local resource quality supports both detrimental and beneficial correlations. The success probability changes along with the fidelity, providing the corresponding yield tradeoff.

SM Sec.~S1 specifies this family, derives a tolerance bound around the independent-input map, and analyzes physical Pauli-$Z$ errors. The physical dephasing example also produces opposite gains at identical single-site and pair error probabilities. Together, the two constructions establish that complete pair characterization leaves the sign of the distillation gain undetermined. The higher-moment map supplies the missing criterion.

Entangled inputs and spin-squeezing-assisted distillation were studied by Bao \emph{et al.}~\cite{bao2022magic}, who found both improved conversion and increased distillation cost for different source states. Their optimized $T$-fidelity selects among Clifford-equivalent target directions. Here independent local twirls produce a separable joint state, and Eq.~(\ref{eq:moment}) evaluates its gain along a fixed target direction. For example, applying these twirls to their GHZ-$T$ family gives
\begin{equation}
 \alpha^2\rho_+^{\otimes5}+\beta^2\rho_-^{\otimes5}
 \quad\Longrightarrow\quad
 p_{\rm acc}=\frac16,\qquad F_{\rm out}=F_{\rm in}=\alpha^2,
 \label{eq:ghzt-twirled-comparison}
\end{equation}
with $\alpha^2+\beta^2=1$. The higher-moment criterion also applies when all single-qubit and two-qubit reduced states remain fixed, as in Fig.~\ref{fig:moment-sensitivity}. At the factory level, O'Gorman and Campbell~\cite{ogorman2017quantum} track correlated output errors and use module-level acceptance to improve repeated distillation. Correlated decoding instead uses the propagation of errors through transversal gates to infer their common origin from joint syndrome records~\cite{cain2024correlated}. The present criterion links joint input moments directly to the accepted resource and provides a local-measurement test of its gain. Stabilizer sampling~\cite{gidney2021stim} and low-rank stabilizer decompositions~\cite{bravyi2019simulation} provide complementary computational approaches for Clifford processing and nonstabilizer inputs, respectively.

\section{Extracting magic from a spin-squeezed precursor}
\label{sec:clifford}
The Ising interaction provides a physical source of the correlations entering this criterion. To identify how its nonstabilizer resource reaches the logical output, consider an isolated block initialized in $|+\rangle^{\otimes5}$ and evolved by $U_{ZZ}=\exp[-i\sum_{i<j}\theta_{ij}Z_iZ_j]$. The global spin flip $L=X^{\otimes5}=\overline X$ commutes with both $U_{ZZ}$ and the code projector. Consequently,
\begin{equation}
 L\Pi U_{ZZ}|+\rangle^{\otimes5}
 =\Pi U_{ZZ}L|+\rangle^{\otimes5}
 =\Pi U_{ZZ}|+\rangle^{\otimes5}.
 \label{eq:parity-proof}
\end{equation}
The accepted state lies in the logical-$X$ $+1$ eigenspace and is therefore $|+\rangle_L$. This conclusion holds for arbitrary weighted Ising couplings. It also covers one-axis twisting, since $J_z^2=5I/4+\tfrac12\sum_{i<j}Z_iZ_j$. The symmetry fixes the projected state even as the interaction produces squeezing and nonstabilizer structure before projection.

An addressed $S_1S_2$ operation changes this selection rule. The transformed parity is $W=(S_1S_2)L(S_1S_2)^\dagger=Y_1Y_2X_3X_4X_5$. It anticommutes with every code generator, so $\Pi W\Pi=0$. The projection therefore removes this parity constraint and can select a nonstabilizer logical state. Exact single-block calculations give a near-unit target fidelity after the logical Clifford alignment $A_L=Y_LH_L$. The Ising interaction supplies the non-Clifford resource; the subsequent phase pattern, code projection, and alignment are stabilizer operations. This is the resource-conversion step linking the squeezed precursor to a usable encoded magic state.

We characterize the precursor using the Wineland parameter $\xi_R^2=5\min_\perp\operatorname{Var}(J_\perp)/|\langle\boldsymbol J\rangle|^2$. The relevant observables are precursor squeezing, accepted logical fidelity, and success probability. They describe different stages of the preparation, and their optima generally occur at different interaction strengths. A uniform $e^{-i\phi J_z}$ rotation provides a comparison source with an additional one-qubit phase resource. SM Sec.~S2 develops the source model. The term spin-squeezed precursor refers to the state before the addressed phase gates, which rotate the individual spin frames.

Interaction-based squeezing also extends to short-range spin exchange~\cite{perlin2020spin} and easy-plane magnetism~\cite{block2024scalable}. These settings motivate evaluating how a chosen code projection converts the prepared correlations into an accepted logical resource.

\section{Geometry controls the correlation contribution}
\label{sec:finite-range}
Large coherent atom arrays~\cite{manetsch2025tweezer} and metasurface-based trapping~\cite{wang2026trapping} motivate layouts that allocate space to both resource blocks and their separation.

\begin{figure*}[!t]
\centering
\includegraphics[width=\textwidth]{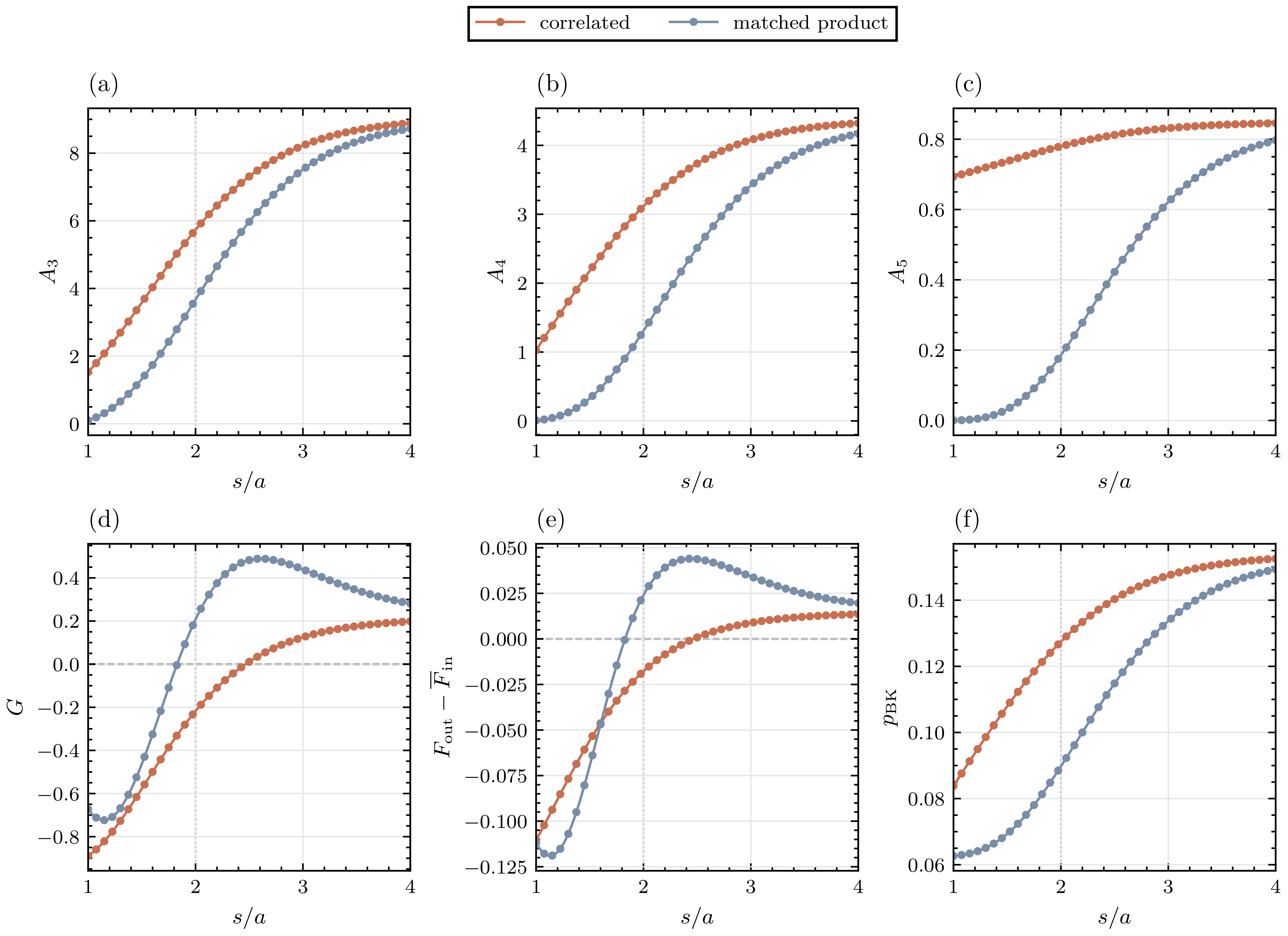}
\caption{Higher moments and distillation gain in the 25-atom source. The source has $(\theta,\phi,R/a)=(0.4,0.3,4)$, with positions $x_{b,j}/a=b(5+s/a)+j$. Red denotes the jointly accepted source after $H_L$ alignment and independent twirls; blue denotes the product of its own five reduced states at each spacing. Panels (a)--(c) show $A_3,A_4,A_5$; (d) shows $G=A_3-2A_5-\bar t(3+A_4)$; (e) shows fidelity gain relative to the mean input; and (f) shows outer success probability. Vertical guides mark the reference spacing near the local-threshold boundary; horizontal zero lines mark the purification boundary. The interior phase slice has a larger local-fidelity margin. Lines join complete code-support evaluations.}
\label{fig:array25-moments}
\end{figure*}

\subsection{Finite-range source and success probability}
We consider five five-atom blocks in a one-dimensional array. Their positions are $x_{b,j}/a=b(5+g)+j$, where $a$ is the intrablock spacing, $g$ is the number of empty guard sites, and $b,j=0,\ldots,4$. The source evolves under
\begin{equation}
 U=\exp\!\left[-\frac{i\theta}{2}\sum_{i<j}
 J_{ij}Z_iZ_j-i\phi J_z\right],
 \label{eq:array}
\end{equation}

\begin{figure}[!t]
\centering
\includegraphics[width=\columnwidth]{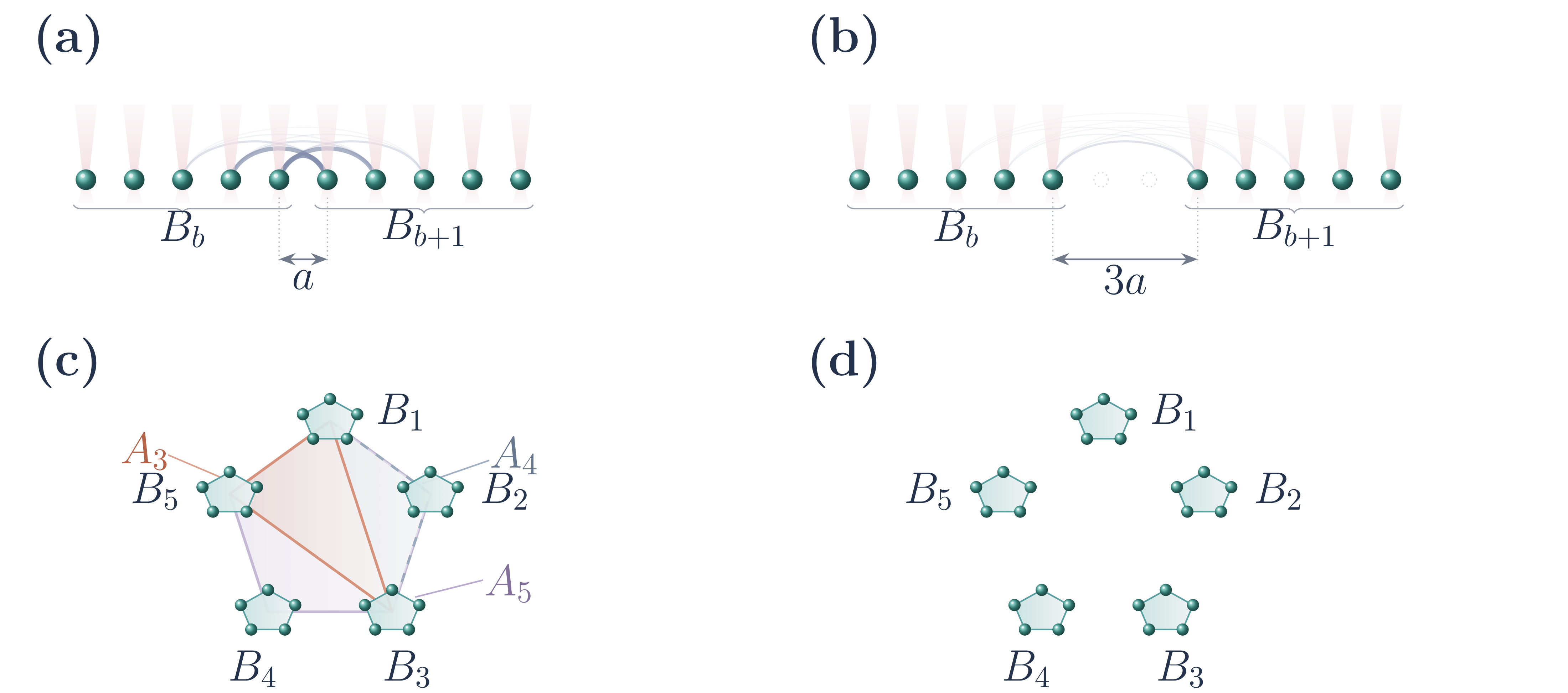}
\caption{Geometry and correlation-sensitive code projection. (a),(b) Two five-atom blocks with zero or two empty guard sites have nearest interblock distances $a$ and $3a$. Coupling-arc contrast follows $J(r)=[1+(r/R)^6]^{-1}$ at $R/a=2.5$. (c),(d) Each pentagon represents a five-atom code block; each panel groups 25 atoms into five blocks. Both inputs have the same twirled reduced logical states $\rho_1,\ldots,\rho_5$. Shaded supports in (c) indicate correlations contributing to $A_3,A_4,A_5$. In (d), the product state's moments factor into local means. Polygon shapes encode the grouping and correlation supports, while the physical model uses a one-dimensional array. The gain comparison is quantified in Fig.~\ref{fig:array25-moments}.}
\label{fig:correlation-mechanism}
\end{figure}

with $J_{ij}=[1+(|x_i-x_j|/R)^6]^{-1}$. The factor $1/2$ matches the one-axis-twisting convention up to a global phase. The addressed source uses the same pair interaction followed by $S_1S_2$ in every block. The uniform-phase source uses the collective rotation shown in Eq.~(\ref{eq:array}). Soft-core interaction models connect this description to Rydberg-dressed squeezing~\cite{young2023enhancing,hines2023spin}.

Spatially tunable interactions~\cite{steinert2023spatially} and control of driven many-body dynamics~\cite{bluvstein2021controlling} make the coupling profile an experimental design variable. Blockade constraints organize collective states~\cite{pan2022composite} and connect scarred dynamics to criticality~\cite{yao2022quantum}; gauge-theory descriptions relate Rydberg models to confinement~\cite{cheng2022tunable} and spin-liquid correlations~\cite{cheng2023variational}. Hybrid analog-digital proposals access nonlocal correlation functions~\cite{li2024uncovering}, while experiments use geometry and system size to control near-critical scaling~\cite{zhang2025observation}. Here, the code projection selects the higher-order moments governing distillation, making block separation a control over their contribution.

We calculate the full jointly accepted logical state, align each logical qubit, and apply independent cyclic Clifford twirls. The calculation retains all computational-basis contributions to the code projection. It therefore propagates the correlations through both the inner preparation and the outer BK step. The total single-shot joint acceptance probability is
\begin{equation}
 p_{\rm tot}=p_{\rm in}^{(5)}\frac{3+A_4}{48}.
 \label{eq:synchronous-yield}
\end{equation}

\begin{figure*}[!t]
\centering
\includegraphics[width=0.95\textwidth]{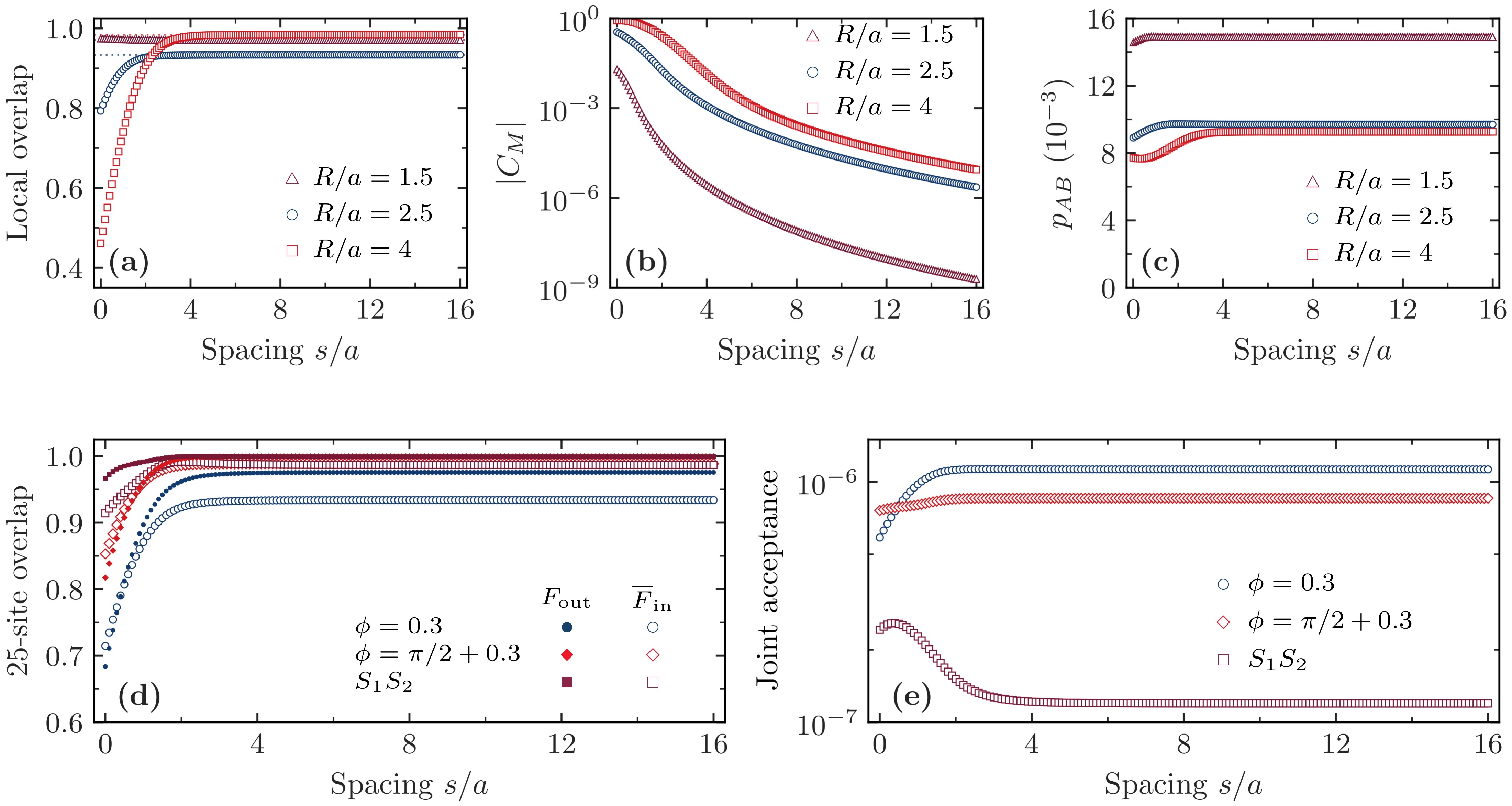}
\caption{Finite-range geometry scan at the reference interaction phase $\theta=0.4$. Continuous spacing $s/a$ agrees with the guard count at integer values. For the two-block uniform-phase source, panels (a)--(c) show local target fidelity, connected covariance magnitude $|C_M|$, and joint inner success probability. The three interaction ranges are given in (a); dotted lines remove interblock couplings. For the five-block source, (d) compares outer output fidelity (filled markers) with mean input fidelity (open markers) for the indicated phase sequences, and (e) gives single-shot inner-plus-outer acceptance. Each point retains the full two-block physical state or five-block code support. Lines connect the evaluated points.}
\label{fig:array}
\end{figure*}

where $p_{\rm in}^{(5)}$ is the joint probability that all five inner projections succeed. SM Sec.~S4 gives the contraction, normalization, and reduced states.

\subsection{A comparison at identical local fidelity}
To separate local-state changes from the effect of correlations, we compare each jointly prepared state with the product of its own five reduced states. Write $t_b=2F_b-1$ and $\bar t=\tfrac15\sum_b t_b$. The product reference has moments
\begin{equation}
 A_k^{\rm prod}=\sum_{|S|=k}\prod_{b\in S}t_b,
 \qquad k=3,4,5.
\end{equation}
This construction preserves every individual input fidelity. Define $G=A_3-2A_5-\bar t(3+A_4)$ and $\Delta A_k=A_k-A_k^{\rm prod}$. The gain and its change relative to the product reference satisfy
\begin{equation}
 \begin{aligned}
 F_{\rm out}-\overline F_{\rm in}
 &=\frac{G}{2(3+A_4)},\\
 G-G^{\rm prod}&=\Delta A_3-2\Delta A_5-\bar t\Delta A_4.
 \end{aligned}
 \label{eq:matched-moment-difference}
\end{equation}
Thus $G$ gives a directly evaluable criterion: its sign determines whether the accepted output improves on the mean input.

Figure~\ref{fig:array25-moments} follows this criterion as whole blocks are translated while their internal lattice remains fixed. The continuous separation $s$ gives positions $x_{b,j}/a=b(5+s/a)+j$ and agrees with the guard geometry when $s/a=g$. The higher-order moments change substantially with separation. In the reversal region, the favorable change in the third-order term is outweighed by the fifth-order term and the fourth-order contribution to normalization. The result is
\begin{equation}
 \min_b F_b>F_{\rm th},\qquad
 F_{\rm out}^{\rm corr}<\overline F,
 \qquad F_{\rm out}^{\rm prod}>\overline F
\label{eq:counterexample-conditions}
\end{equation}
Every input individually exceeds the independent-state threshold, while the correlated joint state loses fidelity and the product of the same local states gains fidelity. Interior points and interaction-strength scans show that this reversal extends across a finite sampled region. The interior phase slice has a finite margin for all three inequalities in Eq.~(\ref{eq:counterexample-conditions}); common drift and position disorder enter through the noise models of SM Sec.~S8~A.

Figure~\ref{fig:correlation-mechanism} relates the moment comparison to the physical controls. Changing the distance between blocks changes the interaction tails and hence the prepared state. The product reference instead isolates the contribution of correlations at fixed reduced states. Using both comparisons connects a physical design choice to the statistical mechanism that determines distillation.

The broader scan in Fig.~\ref{fig:array} shows how this control combines with the addressed Clifford source. Increasing separation suppresses interblock coupling while retaining the interactions within each resource block. The projected state then retains the useful nonstabilizer resource generated by the squeezed precursor and achieves a positive outer distillation gain. Figure~\ref{fig:array}(e) gives the corresponding single-shot joint acceptance probability.

\section{Readout and yield criteria}
\label{sec:readout}
Readout determines which prepared blocks enter the distillation register. Detection and conversion of erasures provide an experimental basis for tracking whether a prepared resource survives~\cite{ma2023high,scholl2023erasure}. We model independent confusion of a complete extracted syndrome bit and independent heralded atom loss. Let $p_0$ and $F_0$ be the ideal inner success probability and true-branch target fidelity, and let $0\leq m\leq1/2$ be the syndrome-bit confusion probability. A reported all-plus outcome can originate from the true sector or a different syndrome sector. These contributions define $h=p_0(1-m)^4$, the false-accept probability $f$, and their sum $r=h+f$.

\begin{figure*}[!t]
\centering
\includegraphics[width=\textwidth]{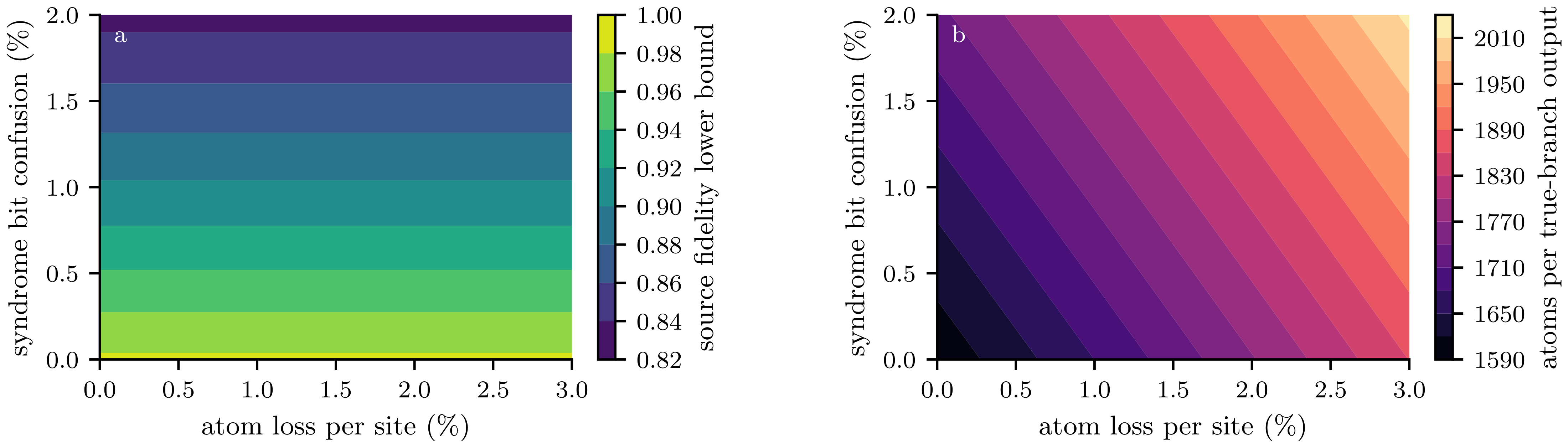}
\caption{Readout and loss sensitivity of the uniform-phase source. (a) Stored-fidelity lower bound for independent syndrome-bit confusion $m$ and arbitrary false-sector outputs. (b) Ideal true-branch loaded-atom cost for independent retries and perfect storage. The panels evaluate Eqs.~(\ref{eq:readout-main}) and (\ref{eq:cost-main}).}
\label{fig:cost}
\end{figure*}

The stored state is the corresponding mixture of true and false accepts. For arbitrary false-sector qubit outputs, $f\leq f_{\max}=(1-p_0)m(1-m)^3$, giving
\begin{equation}
 F_{\rm rep}\geq\frac{hF_0}{h+f_{\max}}
 =\frac{F_0p_0(1-m)}{p_0(1-m)+(1-p_0)m}.
 \label{eq:readout-main}
\end{equation}
Figure~\ref{fig:cost} shows how this bound and the ideal true-branch atom cost respond to report confusion and loss. With independent survival $\kappa=(1-\ell)^5$, the ideal reference cost is
\begin{equation}
 \begin{aligned}
 N_{\rm data}^{\rm ideal}
 &=\frac{25}{\kappa h p_{\rm BK}(\sigma_0^{\otimes5})},\\
 p_{\rm BK}(\sigma_0^{\otimes5})&=\frac{3+5(2F_0-1)^4}{48}.
 \end{aligned}
 \label{eq:cost-main}
\end{equation}
This reference counts true-sector inputs under independent retries and perfect storage. The operational protocol stores every surviving reported accept; its loaded-atom cost is $25/(\kappa r p_{\rm out})$, with $p_{\rm out}$ evaluated on the actual stored-state mixture. SM Sec.~S5 derives the stored-state map, syndrome-resolved bounds, and renewal accounting for repeated outer attempts. Compact ancilla circuits~\cite{chao2018fault,chao2018quantum} offer routes to protecting the check operations entering this accounting.

The syndrome-bit confusion and atom-survival probabilities refer to the complete extraction circuit and preparation shot, respectively.

\section{Source control and experimental diagnosis}
\label{sec:operating-window}
\subsection{Retaining the useful interaction}
The reversal criterion links the microscopic interaction to the source controls. Coupling between blocks changes the joint error distribution entering the outer projection. Within each block, the same interaction creates the precursor from which addressed Clifford phases and code projection extract magic. We retain the intrablock interaction and control its interblock tails through spatial separation. Successful cells are stored and assembled into an outer register. This separates resource generation, correlation diagnosis, and resource delivery within one protocol.

Each source cell undergoes the finite-range interaction, addressed phases, and a six-CZ decoder. An observed report selects a fixed Clifford correction and a keep-or-retry decision. Empty source slots are retried until the register is complete or a finite deadline is reached. The outer circuit acts on the full state conditioned on this common completion history. SM Sec.~S7~B specifies the input frames and report conventions; SM Sec.~S7~C gives the state recurrence and renewal costs.

\begin{figure*}[t]
\centering
\includegraphics[width=0.96\textwidth]{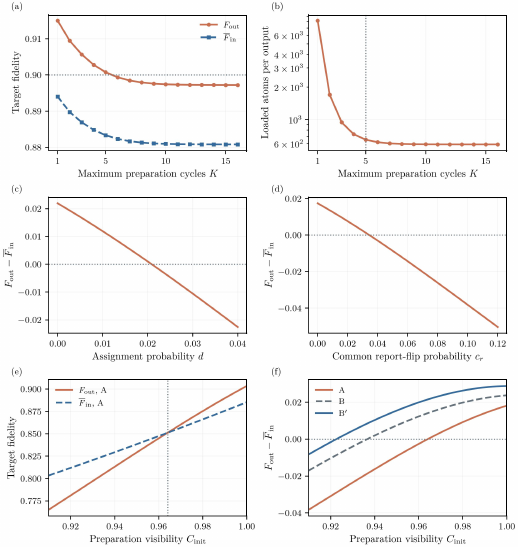}
\caption{Retained-pool operating windows for fixed recovery maps. Panels (a)--(e) use A; panel (f) compares the source settings. (a) Delivered and mean input fidelities versus the preparation deadline. (b) Corresponding loaded-atom cost. (c),(d) Actual outer gain under detector errors and a common report-flip channel. (e) Input and output fidelities for A versus initialized visibility; the vertical guide marks the gain crossing. (f) Visibility dependence of the gain for A, B, and the updated-CZ setting B$^{\prime}$. Every point propagates the joint ready state and its noisy reported instrument. A uses $(\theta,R/a)=(0.4,2.5)$ and B uses $(0.35,4)$. The CZ gate fidelity is $F_2=0.996$ for A and B; B$^{\prime}$ changes it to $0.9985$.}
\label{fig:resource-window}
\end{figure*}

The component benchmarks come from neutral-atom logical processing and CZ gates~\cite{bluvstein2025fault,evered2026nonlocal}, with dressing controls anchored to the spin-squeezing experiment of Ref.~\cite{hines2023spin}. Initialization visibility and accumulated dressing contrast set the quality of the extracted source. At fixed circuit and memory settings, the outer gain becomes positive above the source-contrast threshold. Figure~\ref{fig:resource-window} resolves this boundary together with the fidelity--supply tradeoff. Using the reported initial contrast as the source parameter, the model predicts positive gain for B and the updated-CZ setting B$^{\prime}$ as shown in Fig.~\ref{fig:resource-window}(f). The model separates the effect of local storage from the smaller contribution of common waiting correlations.

For source A, the retained schedule at $K=5$ increases the delivered-output rate and reduces the fresh-data cost relative to synchronous single-shot preparation, while maintaining positive distillation gain. Its deadline balances the supply improvement against dephasing and loss during storage. SM Sec.~S7~E bounds the effect of residual intercell coupling. The calculation charges fresh data atoms to every source attempt. Continuous loading~\cite{chiu2025continuous} and large tweezer arrays~\cite{manetsch2025tweezer,wang2026trapping} supply experimental benchmarks for replenishment and spatial capacity. With refill flux $\Phi_{\rm load}$, the sustainable rate is $\min(\Gamma_{\rm cycle},\Phi_{\rm load}/N_{\rm load})$.

\subsection{Tolerance and a measurable sign test}
The source, extraction circuit, and storage history contribute at different stages. The tolerance analysis varies each model parameter while holding the recovery decisions fixed. A common report flip adds correlated readout errors to the complete inner and outer instruments. A shared logical phase event instead changes the source moments according to $A_k\mapsto[1-c+c(-1/3)^k]A_k$. Independent events with the same local action give a different higher-order response. SM Sec.~S8~A derives the correlated-noise response.

\begin{figure}[t]
\centering
\includegraphics[width=\columnwidth]{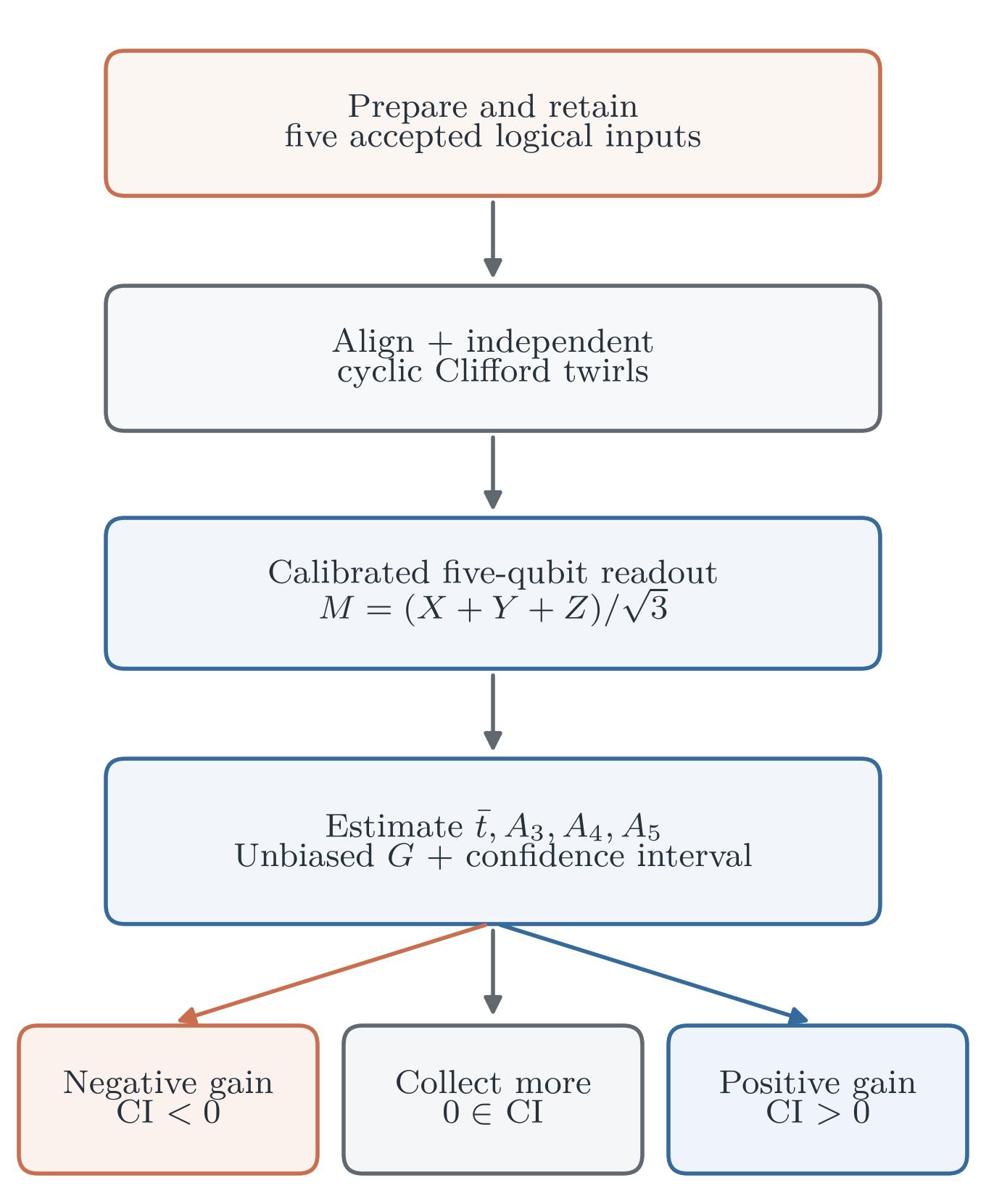}
\caption{Measurement of the correlation criterion. Each accepted five-input register yields the target-axis moments and a confidence interval for $G$. A calibrated response correction includes assignment errors and basis-rotation faults. The sign determines the ideal BK gain; output measurements or a calibrated noisy outer instrument establish the delivered gain.}
\label{fig:diagnostic-flow}
\end{figure}

The criterion $G$ can be measured directly on accepted input registers. After alignment and independent local twirling, target-axis readout yields every required moment from the same record. An unbiased estimator accounts for the product $\bar t A_4$, and calibrated response inversion includes the readout and basis-rotation errors. Distribution-independent and variance-sensitive confidence intervals~\cite{hoeffding1963probability} then determine the sign. Figure~\ref{fig:diagnostic-flow} shows the sequence, and SM Sec.~S9~A derives the finite-sample bounds. The sampling cost scales as $G^{-2}$ near a simple reversal boundary. To test the actual noisy outer gain, the same readout compares independent input registers with delivered outputs.

\subsection{Resource quality and domain of application}
We compare the squeezed source with direct BK preparation at the same delivered quality. The direct input is specified by its post-preparation density matrix; matched ready events impose the same storage ages and downstream channels. Direct preparation uses fewer loaded atoms and avoids the inner decoder when it already provides this source quality. The squeezed route uses the available Ising interaction and Clifford controls to reach that quality through code projection. Its relevant control quantities are the source contrast, the accepted higher moments, and the residual interblock coupling.

Logical BK distillation~\cite{rodriguez2025experimental}, cultivation~\cite{rosenfeld2026cultivation}, and code switching~\cite{daguerre2025experimental} provide complementary experimental resource preparations. Conversion to a common gate-injection interface adds the costs derived in SM Sec.~S6.

The exact moment map follows from the code projector and independent local twirls for any joint input state. The soft-core interaction fixes the numerical source windows, while the specified extraction and storage channels fix the delivered gain and rate. SM Sec.~S8~C bounds residual transverse coherence, and the finite-guard bound controls coupling across the complete preparation history. These relations connect an experimentally measured joint state to its distillation response.

\section{Conclusion}
We have shown that interblock correlations can reverse magic-state distillation while every input remains above the independent-state threshold. The exact five-to-one map separates the correlation orders governing acceptance and output polarization. States with identical one-qubit and two-qubit reduced density matrices can have opposite gains, making the higher-moment criterion $G$ a direct diagnostic of the joint resource.

The spin-squeezed source provides a physical route from this mechanism to control. Addressed Clifford phases remove the parity constraint that otherwise fixes the projected output to a stabilizer state. Spatial separation then controls interblock coupling while preserving the interaction used to generate magic within each block. A retained-cell implementation yields positive noisy gain above a circuit-dependent source-contrast threshold, with the storage deadline and guard spacing setting the quality--supply balance. Target-axis measurements test the correlation criterion, and a calibrated outer instrument connects it to the delivered resource.

\section*{Data availability}
The accompanying executable notebooks contain the numerical models, operating-point data, parameter scans, and figure calculations. The Supplemental Materials derive the projection identities, reported quantum instruments, waiting-history recurrence, and finite-sample statistical bounds.
\begin{acknowledgments}
This work was supported by the National Natural Science Foundation of China (Nos. 12574375 and 12004231), the Fundamental Research Program of Shanxi Province (Grant No. 202203021211301), and the Research Project Supported by Shanxi Scholarship Council of China (Nos. 2023-028 and 2022-014).

We used GPT-6 Astra to assist in generating the schematic illustrations in Figs.~\ref{fig:hierarchical-distillation} and \ref{fig:correlation-mechanism} and in polishing the English text.
\end{acknowledgments}


\begin{thebibliography}{78}%
\makeatletter
\providecommand \@ifxundefined [1]{%
 \@ifx{#1\undefined}
}%
\providecommand \@ifnum [1]{%
 \ifnum #1\expandafter \@firstoftwo
 \else \expandafter \@secondoftwo
 \fi
}%
\providecommand \@ifx [1]{%
 \ifx #1\expandafter \@firstoftwo
 \else \expandafter \@secondoftwo
 \fi
}%
\providecommand \natexlab [1]{#1}%
\providecommand \enquote  [1]{``#1''}%
\providecommand \bibnamefont  [1]{#1}%
\providecommand \bibfnamefont [1]{#1}%
\providecommand \citenamefont [1]{#1}%
\providecommand \href@noop [0]{\@secondoftwo}%
\providecommand \href [0]{\begingroup \@sanitize@url \@href}%
\providecommand \@href[1]{\@@startlink{#1}\@@href}%
\providecommand \@@href[1]{\endgroup#1\@@endlink}%
\providecommand \@sanitize@url [0]{\catcode `\\12\catcode `\$12\catcode
  `\&12\catcode `\#12\catcode `\^12\catcode `\_12\catcode `\%12\relax}%
\providecommand \@@startlink[1]{}%
\providecommand \@@endlink[0]{}%
\providecommand \url  [0]{\begingroup\@sanitize@url \@url }%
\providecommand \@url [1]{\endgroup\@href {#1}{\urlprefix }}%
\providecommand \urlprefix  [0]{URL }%
\providecommand \Eprint [0]{\href }%
\providecommand \doibase [0]{https://doi.org/}%
\providecommand \selectlanguage [0]{\@gobble}%
\providecommand \bibinfo  [0]{\@secondoftwo}%
\providecommand \bibfield  [0]{\@secondoftwo}%
\providecommand \translation [1]{[#1]}%
\providecommand \BibitemOpen [0]{}%
\providecommand \bibitemStop [0]{}%
\providecommand \bibitemNoStop [0]{.\EOS\space}%
\providecommand \EOS [0]{\spacefactor3000\relax}%
\providecommand \BibitemShut  [1]{\csname bibitem#1\endcsname}%
\let\auto@bib@innerbib\@empty
\bibitem [{\citenamefont {Shor}(1995)}]{shor1995scheme}%
  \BibitemOpen
  \bibfield  {author} {\bibinfo {author} {\bibfnamefont {P.~W.}\ \bibnamefont
  {Shor}},\ }\bibfield  {title} {\bibinfo {title} {{Scheme for reducing
  decoherence in quantum computer memory}},\ }\href
  {https://doi.org/10.1103/PhysRevA.52.R2493} {\bibfield  {journal} {\bibinfo
  {journal} {Physical Review A}\ }\textbf {\bibinfo {volume} {52}},\ \bibinfo
  {pages} {R2493} (\bibinfo {year} {1995})}.\BibitemShut {Stop}%
\bibitem [{\citenamefont {Steane}(1996)}]{steane1996error}%
  \BibitemOpen
  \bibfield  {author} {\bibinfo {author} {\bibfnamefont {A.~M.}\ \bibnamefont
  {Steane}},\ }\bibfield  {title} {\bibinfo {title} {{Error Correcting Codes in
  Quantum Theory}},\ }\href {https://doi.org/10.1103/PhysRevLett.77.793}
  {\bibfield  {journal} {\bibinfo  {journal} {Physical Review Letters}\
  }\textbf {\bibinfo {volume} {77}},\ \bibinfo {pages} {793} (\bibinfo {year}
  {1996})}.\BibitemShut {Stop}%
\bibitem [{\citenamefont {Bravyi}\ and\ \citenamefont
  {Kitaev}(2005)}]{bravyi2005universal}%
  \BibitemOpen
  \bibfield  {author} {\bibinfo {author} {\bibfnamefont {S.}~\bibnamefont
  {Bravyi}}\ and\ \bibinfo {author} {\bibfnamefont {A.}~\bibnamefont
  {Kitaev}},\ }\bibfield  {title} {\bibinfo {title} {{Universal quantum
  computation with ideal Clifford gates and noisy ancillas}},\ }\href
  {https://doi.org/10.1103/PhysRevA.71.022316} {\bibfield  {journal} {\bibinfo
  {journal} {Physical Review A}\ }\textbf {\bibinfo {volume} {71}},\ \bibinfo
  {pages} {022316} (\bibinfo {year} {2005})}.\BibitemShut {Stop}%
\bibitem [{\citenamefont {Bravyi}\ and\ \citenamefont
  {Haah}(2012)}]{bravyi2012magic}%
  \BibitemOpen
  \bibfield  {author} {\bibinfo {author} {\bibfnamefont {S.}~\bibnamefont
  {Bravyi}}\ and\ \bibinfo {author} {\bibfnamefont {J.}~\bibnamefont {Haah}},\
  }\bibfield  {title} {\bibinfo {title} {{Magic-state distillation with low
  overhead}},\ }\href {https://doi.org/10.1103/PhysRevA.86.052329} {\bibfield
  {journal} {\bibinfo  {journal} {Physical Review A}\ }\textbf {\bibinfo
  {volume} {86}},\ \bibinfo {pages} {052329} (\bibinfo {year}
  {2012})}.\BibitemShut {Stop}%
\bibitem [{\citenamefont {Jones}(2013)}]{jones2013multilevel}%
  \BibitemOpen
  \bibfield  {author} {\bibinfo {author} {\bibfnamefont {C.}~\bibnamefont
  {Jones}},\ }\bibfield  {title} {\bibinfo {title} {{Multilevel distillation of
  magic states for quantum computing}},\ }\href
  {https://doi.org/10.1103/PhysRevA.87.042305} {\bibfield  {journal} {\bibinfo
  {journal} {Physical Review A}\ }\textbf {\bibinfo {volume} {87}},\ \bibinfo
  {pages} {042305} (\bibinfo {year} {2013})}.\BibitemShut {Stop}%
\bibitem [{\citenamefont {Meier}\ \emph {et~al.}(2013)\citenamefont {Meier},
  \citenamefont {Eastin},\ and\ \citenamefont {Knill}}]{meier2013magic}%
  \BibitemOpen
  \bibfield  {author} {\bibinfo {author} {\bibfnamefont {A.~M.}\ \bibnamefont
  {Meier}}, \bibinfo {author} {\bibfnamefont {B.}~\bibnamefont {Eastin}},\ and\
  \bibinfo {author} {\bibfnamefont {E.~H.}\ \bibnamefont {Knill}},\ }\bibfield
  {title} {\bibinfo {title} {{Magic-state distillation with the four-qubit
  code}},\ }\href {https://doi.org/10.26421/QIC13.3-4-2} {\bibfield  {journal}
  {\bibinfo  {journal} {Quantum Information and Computation}\ }\textbf
  {\bibinfo {volume} {13}},\ \bibinfo {pages} {195} (\bibinfo {year}
  {2013})}.\BibitemShut {Stop}%
\bibitem [{\citenamefont {Campbell}\ and\ \citenamefont
  {Browne}(2010)}]{campbell2010bound}%
  \BibitemOpen
  \bibfield  {author} {\bibinfo {author} {\bibfnamefont {E.~T.}\ \bibnamefont
  {Campbell}}\ and\ \bibinfo {author} {\bibfnamefont {D.~E.}\ \bibnamefont
  {Browne}},\ }\bibfield  {title} {\bibinfo {title} {{Bound States for Magic
  State Distillation in Fault-Tolerant Quantum Computation}},\ }\href
  {https://doi.org/10.1103/PhysRevLett.104.030503} {\bibfield  {journal}
  {\bibinfo  {journal} {Physical Review Letters}\ }\textbf {\bibinfo {volume}
  {104}},\ \bibinfo {pages} {030503} (\bibinfo {year} {2010})}.\BibitemShut {Stop}%
\bibitem [{\citenamefont {Veitch}\ \emph {et~al.}(2014)\citenamefont {Veitch},
  \citenamefont {Mousavian}, \citenamefont {Gottesman},\ and\ \citenamefont
  {Emerson}}]{veitch2014resource}%
  \BibitemOpen
  \bibfield  {author} {\bibinfo {author} {\bibfnamefont {V.}~\bibnamefont
  {Veitch}}, \bibinfo {author} {\bibfnamefont {S.~A.~H.}\ \bibnamefont
  {Mousavian}}, \bibinfo {author} {\bibfnamefont {D.}~\bibnamefont
  {Gottesman}},\ and\ \bibinfo {author} {\bibfnamefont {J.}~\bibnamefont
  {Emerson}},\ }\bibfield  {title} {\bibinfo {title} {{The resource theory of
  stabilizer quantum computation}},\ }\href
  {https://doi.org/10.1088/1367-2630/16/1/013009} {\bibfield  {journal}
  {\bibinfo  {journal} {New Journal of Physics}\ }\textbf {\bibinfo {volume}
  {16}},\ \bibinfo {pages} {013009} (\bibinfo {year} {2014})}.\BibitemShut {Stop}%
\bibitem [{\citenamefont {Howard}\ and\ \citenamefont
  {Campbell}(2017)}]{howard2017application}%
  \BibitemOpen
  \bibfield  {author} {\bibinfo {author} {\bibfnamefont {M.}~\bibnamefont
  {Howard}}\ and\ \bibinfo {author} {\bibfnamefont {E.~T.}\ \bibnamefont
  {Campbell}},\ }\bibfield  {title} {\bibinfo {title} {{Application of a
  Resource Theory for Magic States to Fault-Tolerant Quantum Computing}},\
  }\href {https://doi.org/10.1103/PhysRevLett.118.090501} {\bibfield  {journal}
  {\bibinfo  {journal} {Physical Review Letters}\ }\textbf {\bibinfo {volume}
  {118}},\ \bibinfo {pages} {090501} (\bibinfo {year} {2017})}.\BibitemShut {Stop}%
\bibitem [{\citenamefont {Wang}\ \emph {et~al.}(2020)\citenamefont {Wang},
  \citenamefont {Wilde},\ and\ \citenamefont {Su}}]{wang2020efficiently}%
  \BibitemOpen
  \bibfield  {author} {\bibinfo {author} {\bibfnamefont {X.}~\bibnamefont
  {Wang}}, \bibinfo {author} {\bibfnamefont {M.~M.}\ \bibnamefont {Wilde}},\
  and\ \bibinfo {author} {\bibfnamefont {Y.}~\bibnamefont {Su}},\ }\bibfield
  {title} {\bibinfo {title} {{Efficiently Computable Bounds for Magic State
  Distillation}},\ }\href {https://doi.org/10.1103/PhysRevLett.124.090505}
  {\bibfield  {journal} {\bibinfo  {journal} {Physical Review Letters}\
  }\textbf {\bibinfo {volume} {124}},\ \bibinfo {pages} {090505} (\bibinfo
  {year} {2020})}.\BibitemShut {Stop}%
\bibitem [{\citenamefont {Takagi}\ \emph {et~al.}(2022)\citenamefont {Takagi},
  \citenamefont {Regula},\ and\ \citenamefont {Wilde}}]{takagi2022one}%
  \BibitemOpen
  \bibfield  {author} {\bibinfo {author} {\bibfnamefont {R.}~\bibnamefont
  {Takagi}}, \bibinfo {author} {\bibfnamefont {B.}~\bibnamefont {Regula}},\
  and\ \bibinfo {author} {\bibfnamefont {M.~M.}\ \bibnamefont {Wilde}},\
  }\bibfield  {title} {\bibinfo {title} {{One-Shot Yield-Cost Relations in
  General Quantum Resource Theories}},\ }\href
  {https://doi.org/10.1103/PRXQuantum.3.010348} {\bibfield  {journal} {\bibinfo
   {journal} {PRX Quantum}\ }\textbf {\bibinfo {volume} {3}},\ \bibinfo {pages}
  {010348} (\bibinfo {year} {2022})}.\BibitemShut {Stop}%
\bibitem [{\citenamefont {Litinski}(2019)}]{litinski2019magic}%
  \BibitemOpen
  \bibfield  {author} {\bibinfo {author} {\bibfnamefont {D.}~\bibnamefont
  {Litinski}},\ }\bibfield  {title} {\bibinfo {title} {{Magic State
  Distillation: Not as Costly as You Think}},\ }\href
  {https://doi.org/10.22331/q-2019-12-02-205} {\bibfield  {journal} {\bibinfo
  {journal} {Quantum}\ }\textbf {\bibinfo {volume} {3}},\ \bibinfo {pages}
  {205} (\bibinfo {year} {2019})}.\BibitemShut {Stop}%
\bibitem [{\citenamefont {Gidney}\ and\ \citenamefont
  {Fowler}(2019)}]{gidney2019efficient}%
  \BibitemOpen
  \bibfield  {author} {\bibinfo {author} {\bibfnamefont {C.}~\bibnamefont
  {Gidney}}\ and\ \bibinfo {author} {\bibfnamefont {A.~G.}\ \bibnamefont
  {Fowler}},\ }\bibfield  {title} {\bibinfo {title} {{Efficient magic state
  factories with a catalyzed CCZ-to-2T transformation}},\ }\href
  {https://doi.org/10.22331/q-2019-04-30-135} {\bibfield  {journal} {\bibinfo
  {journal} {Quantum}\ }\textbf {\bibinfo {volume} {3}},\ \bibinfo {pages}
  {135} (\bibinfo {year} {2019})}.\BibitemShut {Stop}%
\bibitem [{\citenamefont {O'Gorman}\ and\ \citenamefont
  {Campbell}(2017)}]{ogorman2017quantum}%
  \BibitemOpen
  \bibfield  {author} {\bibinfo {author} {\bibfnamefont {J.}~\bibnamefont
  {O'Gorman}}\ and\ \bibinfo {author} {\bibfnamefont {E.~T.}\ \bibnamefont
  {Campbell}},\ }\bibfield  {title} {\bibinfo {title} {{Quantum computation
  with realistic magic-state factories}},\ }\href
  {https://doi.org/10.1103/PhysRevA.95.032338} {\bibfield  {journal} {\bibinfo
  {journal} {Physical Review A}\ }\textbf {\bibinfo {volume} {95}},\ \bibinfo
  {pages} {032338} (\bibinfo {year} {2017})}.\BibitemShut {Stop}%
\bibitem [{\citenamefont {Chamberland}\ and\ \citenamefont
  {Cross}(2019)}]{chamberland2019fault}%
  \BibitemOpen
  \bibfield  {author} {\bibinfo {author} {\bibfnamefont {C.}~\bibnamefont
  {Chamberland}}\ and\ \bibinfo {author} {\bibfnamefont {A.~W.}\ \bibnamefont
  {Cross}},\ }\bibfield  {title} {\bibinfo {title} {{Fault-tolerant magic state
  preparation with flag qubits}},\ }\href
  {https://doi.org/10.22331/q-2019-05-20-143} {\bibfield  {journal} {\bibinfo
  {journal} {Quantum}\ }\textbf {\bibinfo {volume} {3}},\ \bibinfo {pages}
  {143} (\bibinfo {year} {2019})}.\BibitemShut {Stop}%
\bibitem [{\citenamefont {Chamberland}\ and\ \citenamefont
  {Noh}(2020)}]{chamberland2020very}%
  \BibitemOpen
  \bibfield  {author} {\bibinfo {author} {\bibfnamefont {C.}~\bibnamefont
  {Chamberland}}\ and\ \bibinfo {author} {\bibfnamefont {K.}~\bibnamefont
  {Noh}},\ }\bibfield  {title} {\bibinfo {title} {{Very low overhead
  fault-tolerant magic state preparation using redundant ancilla encoding and
  flag qubits}},\ }\href {https://doi.org/10.1038/s41534-020-00319-5}
  {\bibfield  {journal} {\bibinfo  {journal} {npj Quantum Information}\
  }\textbf {\bibinfo {volume} {6}},\ \bibinfo {pages} {91} (\bibinfo {year}
  {2020})}.\BibitemShut {Stop}%
\bibitem [{\citenamefont {Bombin}\ and\ \citenamefont
  {Martin-Delgado}(2006)}]{bombin2006topological}%
  \BibitemOpen
  \bibfield  {author} {\bibinfo {author} {\bibfnamefont {H.}~\bibnamefont
  {Bombin}}\ and\ \bibinfo {author} {\bibfnamefont {M.~A.}\ \bibnamefont
  {Martin-Delgado}},\ }\bibfield  {title} {\bibinfo {title} {{Topological
  Quantum Distillation}},\ }\href
  {https://doi.org/10.1103/PhysRevLett.97.180501} {\bibfield  {journal}
  {\bibinfo  {journal} {Physical Review Letters}\ }\textbf {\bibinfo {volume}
  {97}},\ \bibinfo {pages} {180501} (\bibinfo {year} {2006})}.\BibitemShut {Stop}%
\bibitem [{\citenamefont {Kubica}\ \emph {et~al.}(2015)\citenamefont {Kubica},
  \citenamefont {Yoshida},\ and\ \citenamefont
  {Pastawski}}]{kubica2015unfolding}%
  \BibitemOpen
  \bibfield  {author} {\bibinfo {author} {\bibfnamefont {A.}~\bibnamefont
  {Kubica}}, \bibinfo {author} {\bibfnamefont {B.}~\bibnamefont {Yoshida}},\
  and\ \bibinfo {author} {\bibfnamefont {F.}~\bibnamefont {Pastawski}},\
  }\bibfield  {title} {\bibinfo {title} {{Unfolding the color code}},\ }\href
  {https://doi.org/10.1088/1367-2630/17/8/083026} {\bibfield  {journal}
  {\bibinfo  {journal} {New Journal of Physics}\ }\textbf {\bibinfo {volume}
  {17}},\ \bibinfo {pages} {083026} (\bibinfo {year} {2015})}.\BibitemShut {Stop}%
\bibitem [{\citenamefont {Kubica}\ and\ \citenamefont
  {Beverland}(2015)}]{kubica2015universal}%
  \BibitemOpen
  \bibfield  {author} {\bibinfo {author} {\bibfnamefont {A.}~\bibnamefont
  {Kubica}}\ and\ \bibinfo {author} {\bibfnamefont {M.~E.}\ \bibnamefont
  {Beverland}},\ }\bibfield  {title} {\bibinfo {title} {{Universal transversal
  gates with color codes: A simplified approach}},\ }\href
  {https://doi.org/10.1103/PhysRevA.91.032330} {\bibfield  {journal} {\bibinfo
  {journal} {Physical Review A}\ }\textbf {\bibinfo {volume} {91}},\ \bibinfo
  {pages} {032330} (\bibinfo {year} {2015})}.\BibitemShut {Stop}%
\bibitem [{\citenamefont {Bombin}(2015)}]{bombin2015gauge}%
  \BibitemOpen
  \bibfield  {author} {\bibinfo {author} {\bibfnamefont {H.}~\bibnamefont
  {Bombin}},\ }\bibfield  {title} {\bibinfo {title} {{Gauge color codes:
  optimal transversal gates and gauge fixing in topological stabilizer
  codes}},\ }\href {https://doi.org/10.1088/1367-2630/17/8/083002} {\bibfield
  {journal} {\bibinfo  {journal} {New Journal of Physics}\ }\textbf {\bibinfo
  {volume} {17}},\ \bibinfo {pages} {083002} (\bibinfo {year}
  {2015})}.\BibitemShut {Stop}%
\bibitem [{\citenamefont {Anderson}\ \emph {et~al.}(2014)\citenamefont
  {Anderson}, \citenamefont {Duclos-Cianci},\ and\ \citenamefont
  {Poulin}}]{anderson2014fault}%
  \BibitemOpen
  \bibfield  {author} {\bibinfo {author} {\bibfnamefont {J.~T.}\ \bibnamefont
  {Anderson}}, \bibinfo {author} {\bibfnamefont {G.}~\bibnamefont
  {Duclos-Cianci}},\ and\ \bibinfo {author} {\bibfnamefont {D.}~\bibnamefont
  {Poulin}},\ }\bibfield  {title} {\bibinfo {title} {{Fault-Tolerant Conversion
  between the Steane and Reed-Muller Quantum Codes}},\ }\href
  {https://doi.org/10.1103/PhysRevLett.113.080501} {\bibfield  {journal}
  {\bibinfo  {journal} {Physical Review Letters}\ }\textbf {\bibinfo {volume}
  {113}},\ \bibinfo {pages} {080501} (\bibinfo {year} {2014})}.\BibitemShut {Stop}%
\bibitem [{\citenamefont {Beverland}\ \emph {et~al.}(2021)\citenamefont
  {Beverland}, \citenamefont {Kubica},\ and\ \citenamefont
  {Svore}}]{beverland2021cost}%
  \BibitemOpen
  \bibfield  {author} {\bibinfo {author} {\bibfnamefont {M.~E.}\ \bibnamefont
  {Beverland}}, \bibinfo {author} {\bibfnamefont {A.}~\bibnamefont {Kubica}},\
  and\ \bibinfo {author} {\bibfnamefont {K.~M.}\ \bibnamefont {Svore}},\
  }\bibfield  {title} {\bibinfo {title} {{Cost of Universality: A Comparative
  Study of the Overhead of State Distillation and Code Switching with Color
  Codes}},\ }\href {https://doi.org/10.1103/PRXQuantum.2.020341} {\bibfield
  {journal} {\bibinfo  {journal} {PRX Quantum}\ }\textbf {\bibinfo {volume}
  {2}},\ \bibinfo {pages} {020341} (\bibinfo {year} {2021})}.\BibitemShut {Stop}%
\bibitem [{\citenamefont {Bao}\ \emph {et~al.}(2022)\citenamefont {Bao},
  \citenamefont {Cao},\ and\ \citenamefont {Su}}]{bao2022magic}%
  \BibitemOpen
  \bibfield  {author} {\bibinfo {author} {\bibfnamefont {N.}~\bibnamefont
  {Bao}}, \bibinfo {author} {\bibfnamefont {C.}~\bibnamefont {Cao}},\ and\
  \bibinfo {author} {\bibfnamefont {V.~P.}\ \bibnamefont {Su}},\ }\bibfield
  {title} {\bibinfo {title} {{Magic state distillation from entangled
  states}},\ }\href {https://doi.org/10.1103/PhysRevA.105.022602} {\bibfield
  {journal} {\bibinfo  {journal} {Physical Review A}\ }\textbf {\bibinfo
  {volume} {105}},\ \bibinfo {pages} {022602} (\bibinfo {year}
  {2022})}.\BibitemShut {Stop}%
\bibitem [{\citenamefont {Kitagawa}\ and\ \citenamefont
  {Ueda}(1993)}]{kitagawa1993squeezed}%
  \BibitemOpen
  \bibfield  {author} {\bibinfo {author} {\bibfnamefont {M.}~\bibnamefont
  {Kitagawa}}\ and\ \bibinfo {author} {\bibfnamefont {M.}~\bibnamefont
  {Ueda}},\ }\bibfield  {title} {\bibinfo {title} {{Squeezed spin states}},\
  }\href {https://doi.org/10.1103/PhysRevA.47.5138} {\bibfield  {journal}
  {\bibinfo  {journal} {Physical Review A}\ }\textbf {\bibinfo {volume} {47}},\
  \bibinfo {pages} {5138} (\bibinfo {year} {1993})}.\BibitemShut {Stop}%
\bibitem [{\citenamefont {Toth}\ \emph {et~al.}(2009)\citenamefont {Toth},
  \citenamefont {Knapp}, \citenamefont {Guhne},\ and\ \citenamefont
  {Briegel}}]{toth2009spin}%
  \BibitemOpen
  \bibfield  {author} {\bibinfo {author} {\bibfnamefont {G.}~\bibnamefont
  {Toth}}, \bibinfo {author} {\bibfnamefont {C.}~\bibnamefont {Knapp}},
  \bibinfo {author} {\bibfnamefont {O.}~\bibnamefont {Guhne}},\ and\ \bibinfo
  {author} {\bibfnamefont {H.~J.}\ \bibnamefont {Briegel}},\ }\bibfield
  {title} {\bibinfo {title} {{Spin squeezing and entanglement}},\ }\href
  {https://doi.org/10.1103/PhysRevA.79.042334} {\bibfield  {journal} {\bibinfo
  {journal} {Physical Review A}\ }\textbf {\bibinfo {volume} {79}},\ \bibinfo
  {pages} {042334} (\bibinfo {year} {2009})}.\BibitemShut {Stop}%
\bibitem [{\citenamefont {Leroux}\ \emph {et~al.}(2010)\citenamefont {Leroux},
  \citenamefont {Schleier-Smith},\ and\ \citenamefont
  {Vuletic}}]{leroux2010implementation}%
  \BibitemOpen
  \bibfield  {author} {\bibinfo {author} {\bibfnamefont {I.~D.}\ \bibnamefont
  {Leroux}}, \bibinfo {author} {\bibfnamefont {M.}~\bibnamefont
  {Schleier-Smith}},\ and\ \bibinfo {author} {\bibfnamefont {V.}~\bibnamefont
  {Vuletic}},\ }\bibfield  {title} {\bibinfo {title} {{Implementation of Cavity
  Squeezing of a Collective Atomic Spin}},\ }\href
  {https://doi.org/10.1103/PhysRevLett.104.073602} {\bibfield  {journal}
  {\bibinfo  {journal} {Physical Review Letters}\ }\textbf {\bibinfo {volume}
  {104}},\ \bibinfo {pages} {073602} (\bibinfo {year} {2010})}.\BibitemShut {Stop}%
\bibitem [{\citenamefont {Schleier-Smith}\ \emph {et~al.}(2010)\citenamefont
  {Schleier-Smith}, \citenamefont {Leroux},\ and\ \citenamefont
  {Vuletic}}]{schleiersmith2010squeezing}%
  \BibitemOpen
  \bibfield  {author} {\bibinfo {author} {\bibfnamefont {M.}~\bibnamefont
  {Schleier-Smith}}, \bibinfo {author} {\bibfnamefont {I.~D.}\ \bibnamefont
  {Leroux}},\ and\ \bibinfo {author} {\bibfnamefont {V.}~\bibnamefont
  {Vuletic}},\ }\bibfield  {title} {\bibinfo {title} {{Squeezing the collective
  spin of a dilute atomic ensemble by cavity feedback}},\ }\href
  {https://doi.org/10.1103/PhysRevA.81.021804} {\bibfield  {journal} {\bibinfo
  {journal} {Physical Review A}\ }\textbf {\bibinfo {volume} {81}},\ \bibinfo
  {pages} {021804} (\bibinfo {year} {2010})}.\BibitemShut {Stop}%
\bibitem [{\citenamefont {Bornet}\ \emph {et~al.}(2023)\citenamefont {Bornet},
  \citenamefont {Emperauger}, \citenamefont {Chen}, \citenamefont {Ye},
  \citenamefont {Block}, \citenamefont {Bintz} \emph
  {et~al.}}]{bornet2023scalable}%
  \BibitemOpen
  \bibfield  {author} {\bibinfo {author} {\bibfnamefont {G.}~\bibnamefont
  {Bornet}}, \bibinfo {author} {\bibfnamefont {G.}~\bibnamefont {Emperauger}},
  \bibinfo {author} {\bibfnamefont {C.}~\bibnamefont {Chen}}, \bibinfo {author}
  {\bibfnamefont {B.}~\bibnamefont {Ye}}, \bibinfo {author} {\bibfnamefont
  {M.}~\bibnamefont {Block}}, \bibinfo {author} {\bibfnamefont
  {M.}~\bibnamefont {Bintz}}, \emph {et~al.},\ }\bibfield  {title} {\bibinfo
  {title} {{Scalable spin squeezing in a dipolar Rydberg atom array}},\ }\href
  {https://doi.org/10.1038/s41586-023-06414-9} {\bibfield  {journal} {\bibinfo
  {journal} {Nature}\ }\textbf {\bibinfo {volume} {621}},\ \bibinfo {pages}
  {728} (\bibinfo {year} {2023})}.\BibitemShut {Stop}%
\bibitem [{\citenamefont {Eckner}\ \emph {et~al.}(2023)\citenamefont {Eckner},
  \citenamefont {Oppong}, \citenamefont {Cao}, \citenamefont {Young},
  \citenamefont {Milner}, \citenamefont {Robinson}, \citenamefont {Ye},\ and\
  \citenamefont {Kaufman}}]{eckner2023realizing}%
  \BibitemOpen
  \bibfield  {author} {\bibinfo {author} {\bibfnamefont {W.~J.}\ \bibnamefont
  {Eckner}}, \bibinfo {author} {\bibfnamefont {N.~D.}\ \bibnamefont {Oppong}},
  \bibinfo {author} {\bibfnamefont {A.}~\bibnamefont {Cao}}, \bibinfo {author}
  {\bibfnamefont {A.~W.}\ \bibnamefont {Young}}, \bibinfo {author}
  {\bibfnamefont {W.~R.}\ \bibnamefont {Milner}}, \bibinfo {author}
  {\bibfnamefont {J.~M.}\ \bibnamefont {Robinson}}, \bibinfo {author}
  {\bibfnamefont {J.}~\bibnamefont {Ye}},\ and\ \bibinfo {author}
  {\bibfnamefont {A.~M.}\ \bibnamefont {Kaufman}},\ }\bibfield  {title}
  {\bibinfo {title} {{Realizing spin squeezing with Rydberg interactions in an
  optical clock}},\ }\href {https://doi.org/10.1038/s41586-023-06360-6}
  {\bibfield  {journal} {\bibinfo  {journal} {Nature}\ }\textbf {\bibinfo
  {volume} {621}},\ \bibinfo {pages} {734} (\bibinfo {year}
  {2023})}.\BibitemShut {Stop}%
\bibitem [{\citenamefont {Hines}\ \emph {et~al.}(2023)\citenamefont {Hines},
  \citenamefont {Rajagopal}, \citenamefont {Moreau}, \citenamefont {Wahrman},
  \citenamefont {Lewis}, \citenamefont {Markovic},\ and\ \citenamefont
  {Schleier-Smith}}]{hines2023spin}%
  \BibitemOpen
  \bibfield  {author} {\bibinfo {author} {\bibfnamefont {J.}~\bibnamefont
  {Hines}}, \bibinfo {author} {\bibfnamefont {S.}~\bibnamefont {Rajagopal}},
  \bibinfo {author} {\bibfnamefont {G.~L.}\ \bibnamefont {Moreau}}, \bibinfo
  {author} {\bibfnamefont {M.}~\bibnamefont {Wahrman}}, \bibinfo {author}
  {\bibfnamefont {N.~A.}\ \bibnamefont {Lewis}}, \bibinfo {author}
  {\bibfnamefont {O.}~\bibnamefont {Markovic}},\ and\ \bibinfo {author}
  {\bibfnamefont {M.}~\bibnamefont {Schleier-Smith}},\ }\bibfield  {title}
  {\bibinfo {title} {{Spin Squeezing by Rydberg Dressing in an Array of Atomic
  Ensembles}},\ }\href {https://doi.org/10.1103/PhysRevLett.131.063401}
  {\bibfield  {journal} {\bibinfo  {journal} {Physical Review Letters}\
  }\textbf {\bibinfo {volume} {131}},\ \bibinfo {pages} {063401} (\bibinfo
  {year} {2023})}.\BibitemShut {Stop}%
\bibitem [{\citenamefont {Souza}\ \emph {et~al.}(2011)\citenamefont {Souza},
  \citenamefont {Zhang}, \citenamefont {Ryan},\ and\ \citenamefont
  {Laflamme}}]{souza2011experimental}%
  \BibitemOpen
  \bibfield  {author} {\bibinfo {author} {\bibfnamefont {A.~M.}\ \bibnamefont
  {Souza}}, \bibinfo {author} {\bibfnamefont {J.}~\bibnamefont {Zhang}},
  \bibinfo {author} {\bibfnamefont {C.~A.}\ \bibnamefont {Ryan}},\ and\
  \bibinfo {author} {\bibfnamefont {R.}~\bibnamefont {Laflamme}},\ }\bibfield
  {title} {\bibinfo {title} {{Experimental magic state distillation for
  fault-tolerant quantum computing}},\ }\href
  {https://doi.org/10.1038/ncomms1166} {\bibfield  {journal} {\bibinfo
  {journal} {Nature Communications}\ }\textbf {\bibinfo {volume} {2}},\
  \bibinfo {pages} {169} (\bibinfo {year} {2011})}.\BibitemShut {Stop}%
\bibitem [{\citenamefont {Postler}\ \emph {et~al.}(2022)\citenamefont
  {Postler}, \citenamefont {Heu{\ss}en}, \citenamefont {Pogorelov},
  \citenamefont {Rispler}, \citenamefont {Feldker}, \citenamefont {Meth} \emph
  {et~al.}}]{postler2022demonstration}%
  \BibitemOpen
  \bibfield  {author} {\bibinfo {author} {\bibfnamefont {L.}~\bibnamefont
  {Postler}}, \bibinfo {author} {\bibfnamefont {S.}~\bibnamefont {Heu{\ss}en}},
  \bibinfo {author} {\bibfnamefont {I.}~\bibnamefont {Pogorelov}}, \bibinfo
  {author} {\bibfnamefont {M.}~\bibnamefont {Rispler}}, \bibinfo {author}
  {\bibfnamefont {T.}~\bibnamefont {Feldker}}, \bibinfo {author} {\bibfnamefont
  {M.}~\bibnamefont {Meth}}, \emph {et~al.},\ }\bibfield  {title} {\bibinfo
  {title} {{Demonstration of fault-tolerant universal quantum gate
  operations}},\ }\href {https://doi.org/10.1038/s41586-022-04721-1} {\bibfield
   {journal} {\bibinfo  {journal} {Nature}\ }\textbf {\bibinfo {volume}
  {605}},\ \bibinfo {pages} {675} (\bibinfo {year} {2022})}.\BibitemShut {Stop}%
\bibitem [{\citenamefont {Ye}\ \emph {et~al.}(2023)\citenamefont {Ye},
  \citenamefont {He}, \citenamefont {Huang}, \citenamefont {Wei}, \citenamefont
  {Zhang}, \citenamefont {Zhao} \emph {et~al.}}]{ye2023logical}%
  \BibitemOpen
  \bibfield  {author} {\bibinfo {author} {\bibfnamefont {Y.}~\bibnamefont
  {Ye}}, \bibinfo {author} {\bibfnamefont {T.}~\bibnamefont {He}}, \bibinfo
  {author} {\bibfnamefont {H.-L.}\ \bibnamefont {Huang}}, \bibinfo {author}
  {\bibfnamefont {Z.}~\bibnamefont {Wei}}, \bibinfo {author} {\bibfnamefont
  {Y.}~\bibnamefont {Zhang}}, \bibinfo {author} {\bibfnamefont
  {Y.}~\bibnamefont {Zhao}}, \emph {et~al.},\ }\bibfield  {title} {\bibinfo
  {title} {{Logical Magic State Preparation with Fidelity beyond the
  Distillation Threshold on a Superconducting Quantum Processor}},\ }\href
  {https://doi.org/10.1103/PhysRevLett.131.210603} {\bibfield  {journal}
  {\bibinfo  {journal} {Physical Review Letters}\ }\textbf {\bibinfo {volume}
  {131}},\ \bibinfo {pages} {210603} (\bibinfo {year} {2023})}.\BibitemShut {Stop}%
\bibitem [{\citenamefont {Gupta}\ \emph {et~al.}(2024)\citenamefont {Gupta},
  \citenamefont {Sundaresan}, \citenamefont {Alexander}, \citenamefont {Wood},
  \citenamefont {Merkel}, \citenamefont {Healy} \emph
  {et~al.}}]{gupta2024encoding}%
  \BibitemOpen
  \bibfield  {author} {\bibinfo {author} {\bibfnamefont {R.~S.}\ \bibnamefont
  {Gupta}}, \bibinfo {author} {\bibfnamefont {N.}~\bibnamefont {Sundaresan}},
  \bibinfo {author} {\bibfnamefont {T.~J.}\ \bibnamefont {Alexander}}, \bibinfo
  {author} {\bibfnamefont {C.~J.}\ \bibnamefont {Wood}}, \bibinfo {author}
  {\bibfnamefont {S.}~\bibnamefont {Merkel}}, \bibinfo {author} {\bibfnamefont
  {M.~B.}\ \bibnamefont {Healy}}, \emph {et~al.},\ }\bibfield  {title}
  {\bibinfo {title} {{Encoding a magic state with beyond break-even
  fidelity}},\ }\href {https://doi.org/10.1038/s41586-023-06846-3} {\bibfield
  {journal} {\bibinfo  {journal} {Nature}\ }\textbf {\bibinfo {volume} {625}},\
  \bibinfo {pages} {259} (\bibinfo {year} {2024})}.\BibitemShut {Stop}%
\bibitem [{\citenamefont {{Google Quantum AI and
  Collaborators}}(2025)}]{google2025below}%
  \BibitemOpen
  \bibfield  {author} {\bibinfo {author} {\bibnamefont {{Google Quantum AI and
  Collaborators}}},\ }\bibfield  {title} {\bibinfo {title} {{Quantum error
  correction below the surface code threshold}},\ }\href
  {https://doi.org/10.1038/s41586-024-08449-y} {\bibfield  {journal} {\bibinfo
  {journal} {Nature}\ }\textbf {\bibinfo {volume} {638}},\ \bibinfo {pages}
  {920} (\bibinfo {year} {2025})}.\BibitemShut {Stop}%
\bibitem [{\citenamefont {Rodriguez}\ \emph {et~al.}(2025)\citenamefont
  {Rodriguez}, \citenamefont {Robinson}, \citenamefont {Jepsen}, \citenamefont
  {He}, \citenamefont {Duckering}, \citenamefont {Zhao} \emph
  {et~al.}}]{rodriguez2025experimental}%
  \BibitemOpen
  \bibfield  {author} {\bibinfo {author} {\bibfnamefont {P.~S.}\ \bibnamefont
  {Rodriguez}}, \bibinfo {author} {\bibfnamefont {J.~M.}\ \bibnamefont
  {Robinson}}, \bibinfo {author} {\bibfnamefont {P.~N.}\ \bibnamefont
  {Jepsen}}, \bibinfo {author} {\bibfnamefont {Z.}~\bibnamefont {He}}, \bibinfo
  {author} {\bibfnamefont {C.}~\bibnamefont {Duckering}}, \bibinfo {author}
  {\bibfnamefont {C.}~\bibnamefont {Zhao}}, \emph {et~al.},\ }\bibfield
  {title} {\bibinfo {title} {{Experimental demonstration of logical magic state
  distillation}},\ }\href {https://doi.org/10.1038/s41586-025-09367-3}
  {\bibfield  {journal} {\bibinfo  {journal} {Nature}\ }\textbf {\bibinfo
  {volume} {645}},\ \bibinfo {pages} {620} (\bibinfo {year}
  {2025})}.\BibitemShut {Stop}%
\bibitem [{\citenamefont {Rosenfeld}\ \emph {et~al.}(2026)\citenamefont
  {Rosenfeld}, \citenamefont {Gidney}, \citenamefont {Roberts}, \citenamefont
  {Morvan}, \citenamefont {Lacroix}, \citenamefont {Kafri} \emph
  {et~al.}}]{rosenfeld2026cultivation}%
  \BibitemOpen
  \bibfield  {author} {\bibinfo {author} {\bibfnamefont {E.}~\bibnamefont
  {Rosenfeld}}, \bibinfo {author} {\bibfnamefont {C.}~\bibnamefont {Gidney}},
  \bibinfo {author} {\bibfnamefont {G.}~\bibnamefont {Roberts}}, \bibinfo
  {author} {\bibfnamefont {A.}~\bibnamefont {Morvan}}, \bibinfo {author}
  {\bibfnamefont {N.}~\bibnamefont {Lacroix}}, \bibinfo {author} {\bibfnamefont
  {D.}~\bibnamefont {Kafri}}, \emph {et~al.},\ }\bibfield  {title} {\bibinfo
  {title} {{Magic state cultivation on a superconducting quantum processor}},\
  }\href {https://doi.org/10.1126/science.aee7961} {\bibfield  {journal}
  {\bibinfo  {journal} {Science}\ ,\ \bibinfo {pages} {eaee7961}} (\bibinfo
  {year} {2026})}.\BibitemShut {Stop}%
\bibitem [{\citenamefont {Barredo}\ \emph {et~al.}(2016)\citenamefont
  {Barredo}, \citenamefont {de~Leseleuc}, \citenamefont {Lienhard},
  \citenamefont {Lahaye},\ and\ \citenamefont {Browaeys}}]{barredo2016atom}%
  \BibitemOpen
  \bibfield  {author} {\bibinfo {author} {\bibfnamefont {D.}~\bibnamefont
  {Barredo}}, \bibinfo {author} {\bibfnamefont {S.}~\bibnamefont
  {de~Leseleuc}}, \bibinfo {author} {\bibfnamefont {V.}~\bibnamefont
  {Lienhard}}, \bibinfo {author} {\bibfnamefont {T.}~\bibnamefont {Lahaye}},\
  and\ \bibinfo {author} {\bibfnamefont {A.}~\bibnamefont {Browaeys}},\
  }\bibfield  {title} {\bibinfo {title} {{An atom-by-atom assembler of
  defect-free arbitrary two-dimensional atomic arrays}},\ }\href
  {https://doi.org/10.1126/science.aah3778} {\bibfield  {journal} {\bibinfo
  {journal} {Science}\ }\textbf {\bibinfo {volume} {354}},\ \bibinfo {pages}
  {1021} (\bibinfo {year} {2016})}.\BibitemShut {Stop}%
\bibitem [{\citenamefont {Endres}\ \emph {et~al.}(2016)\citenamefont {Endres},
  \citenamefont {Bernien}, \citenamefont {Keesling}, \citenamefont {Levine},
  \citenamefont {Anschuetz}, \citenamefont {Krajenbrink}, \citenamefont
  {Senko}, \citenamefont {Vuletic}, \citenamefont {Greiner},\ and\
  \citenamefont {Lukin}}]{endres2016atom}%
  \BibitemOpen
  \bibfield  {author} {\bibinfo {author} {\bibfnamefont {M.}~\bibnamefont
  {Endres}}, \bibinfo {author} {\bibfnamefont {H.}~\bibnamefont {Bernien}},
  \bibinfo {author} {\bibfnamefont {A.}~\bibnamefont {Keesling}}, \bibinfo
  {author} {\bibfnamefont {H.}~\bibnamefont {Levine}}, \bibinfo {author}
  {\bibfnamefont {E.~R.}\ \bibnamefont {Anschuetz}}, \bibinfo {author}
  {\bibfnamefont {A.}~\bibnamefont {Krajenbrink}}, \bibinfo {author}
  {\bibfnamefont {C.}~\bibnamefont {Senko}}, \bibinfo {author} {\bibfnamefont
  {V.}~\bibnamefont {Vuletic}}, \bibinfo {author} {\bibfnamefont
  {M.}~\bibnamefont {Greiner}},\ and\ \bibinfo {author} {\bibfnamefont {M.~D.}\
  \bibnamefont {Lukin}},\ }\bibfield  {title} {\bibinfo {title} {{Atom-by-atom
  assembly of defect-free one-dimensional cold atom arrays}},\ }\href
  {https://doi.org/10.1126/science.aah3752} {\bibfield  {journal} {\bibinfo
  {journal} {Science}\ }\textbf {\bibinfo {volume} {354}},\ \bibinfo {pages}
  {1024} (\bibinfo {year} {2016})}.\BibitemShut {Stop}%
\bibitem [{\citenamefont {Ebadi}\ \emph {et~al.}(2021)\citenamefont {Ebadi},
  \citenamefont {Wang}, \citenamefont {Levine}, \citenamefont {Keesling},
  \citenamefont {Semeghini}, \citenamefont {Omran} \emph
  {et~al.}}]{ebadi2021quantum}%
  \BibitemOpen
  \bibfield  {author} {\bibinfo {author} {\bibfnamefont {S.}~\bibnamefont
  {Ebadi}}, \bibinfo {author} {\bibfnamefont {T.~T.}\ \bibnamefont {Wang}},
  \bibinfo {author} {\bibfnamefont {H.}~\bibnamefont {Levine}}, \bibinfo
  {author} {\bibfnamefont {A.}~\bibnamefont {Keesling}}, \bibinfo {author}
  {\bibfnamefont {G.}~\bibnamefont {Semeghini}}, \bibinfo {author}
  {\bibfnamefont {A.}~\bibnamefont {Omran}}, \emph {et~al.},\ }\bibfield
  {title} {\bibinfo {title} {{Quantum phases of matter on a 256-atom
  programmable quantum simulator}},\ }\href
  {https://doi.org/10.1038/s41586-021-03582-4} {\bibfield  {journal} {\bibinfo
  {journal} {Nature}\ }\textbf {\bibinfo {volume} {595}},\ \bibinfo {pages}
  {227} (\bibinfo {year} {2021})}.\BibitemShut {Stop}%
\bibitem [{\citenamefont {Bluvstein}\ \emph {et~al.}(2022)\citenamefont
  {Bluvstein}, \citenamefont {Levine}, \citenamefont {Semeghini}, \citenamefont
  {Wang}, \citenamefont {Ebadi}, \citenamefont {Kalinowski} \emph
  {et~al.}}]{bluvstein2022quantum}%
  \BibitemOpen
  \bibfield  {author} {\bibinfo {author} {\bibfnamefont {D.}~\bibnamefont
  {Bluvstein}}, \bibinfo {author} {\bibfnamefont {H.}~\bibnamefont {Levine}},
  \bibinfo {author} {\bibfnamefont {G.}~\bibnamefont {Semeghini}}, \bibinfo
  {author} {\bibfnamefont {T.~T.}\ \bibnamefont {Wang}}, \bibinfo {author}
  {\bibfnamefont {S.}~\bibnamefont {Ebadi}}, \bibinfo {author} {\bibfnamefont
  {M.~W.}\ \bibnamefont {Kalinowski}}, \emph {et~al.},\ }\bibfield  {title}
  {\bibinfo {title} {{A quantum processor based on coherent transport of
  entangled atom arrays}},\ }\href {https://doi.org/10.1038/s41586-022-04592-6}
  {\bibfield  {journal} {\bibinfo  {journal} {Nature}\ }\textbf {\bibinfo
  {volume} {604}},\ \bibinfo {pages} {451} (\bibinfo {year}
  {2022})}.\BibitemShut {Stop}%
\bibitem [{\citenamefont {Bluvstein}\ \emph {et~al.}(2024)\citenamefont
  {Bluvstein}, \citenamefont {Evered}, \citenamefont {Geim}, \citenamefont
  {Li}, \citenamefont {Zhou}, \citenamefont {Manovitz} \emph
  {et~al.}}]{bluvstein2023logical}%
  \BibitemOpen
  \bibfield  {author} {\bibinfo {author} {\bibfnamefont {D.}~\bibnamefont
  {Bluvstein}}, \bibinfo {author} {\bibfnamefont {S.~J.}\ \bibnamefont
  {Evered}}, \bibinfo {author} {\bibfnamefont {A.~A.}\ \bibnamefont {Geim}},
  \bibinfo {author} {\bibfnamefont {S.~H.}\ \bibnamefont {Li}}, \bibinfo
  {author} {\bibfnamefont {H.}~\bibnamefont {Zhou}}, \bibinfo {author}
  {\bibfnamefont {T.}~\bibnamefont {Manovitz}}, \emph {et~al.},\ }\bibfield
  {title} {\bibinfo {title} {{Logical quantum processor based on reconfigurable
  atom arrays}},\ }\href {https://doi.org/10.1038/s41586-023-06927-3}
  {\bibfield  {journal} {\bibinfo  {journal} {Nature}\ }\textbf {\bibinfo
  {volume} {626}},\ \bibinfo {pages} {58} (\bibinfo {year} {2024})}.\BibitemShut {Stop}%
\bibitem [{\citenamefont {Levine}\ \emph {et~al.}(2022)\citenamefont {Levine},
  \citenamefont {Bluvstein}, \citenamefont {Keesling}, \citenamefont {Wang},
  \citenamefont {Ebadi}, \citenamefont {Semeghini}, \citenamefont {Omran},
  \citenamefont {Greiner}, \citenamefont {Vuletic},\ and\ \citenamefont
  {Lukin}}]{levine2022dispersive}%
  \BibitemOpen
  \bibfield  {author} {\bibinfo {author} {\bibfnamefont {H.}~\bibnamefont
  {Levine}}, \bibinfo {author} {\bibfnamefont {D.}~\bibnamefont {Bluvstein}},
  \bibinfo {author} {\bibfnamefont {A.}~\bibnamefont {Keesling}}, \bibinfo
  {author} {\bibfnamefont {T.~T.}\ \bibnamefont {Wang}}, \bibinfo {author}
  {\bibfnamefont {S.}~\bibnamefont {Ebadi}}, \bibinfo {author} {\bibfnamefont
  {G.}~\bibnamefont {Semeghini}}, \bibinfo {author} {\bibfnamefont
  {A.}~\bibnamefont {Omran}}, \bibinfo {author} {\bibfnamefont
  {M.}~\bibnamefont {Greiner}}, \bibinfo {author} {\bibfnamefont
  {V.}~\bibnamefont {Vuletic}},\ and\ \bibinfo {author} {\bibfnamefont {M.~D.}\
  \bibnamefont {Lukin}},\ }\bibfield  {title} {\bibinfo {title} {{Dispersive
  optical systems for scalable Raman driving of hyperfine qubits}},\ }\href
  {https://doi.org/10.1103/PhysRevA.105.032618} {\bibfield  {journal} {\bibinfo
   {journal} {Physical Review A}\ }\textbf {\bibinfo {volume} {105}},\ \bibinfo
  {pages} {032618} (\bibinfo {year} {2022})}.\BibitemShut {Stop}%
\bibitem [{\citenamefont {Levine}\ \emph {et~al.}(2019)\citenamefont {Levine},
  \citenamefont {Keesling}, \citenamefont {Semeghini}, \citenamefont {Omran},
  \citenamefont {Wang}, \citenamefont {Ebadi} \emph
  {et~al.}}]{levine2019parallel}%
  \BibitemOpen
  \bibfield  {author} {\bibinfo {author} {\bibfnamefont {H.}~\bibnamefont
  {Levine}}, \bibinfo {author} {\bibfnamefont {A.}~\bibnamefont {Keesling}},
  \bibinfo {author} {\bibfnamefont {G.}~\bibnamefont {Semeghini}}, \bibinfo
  {author} {\bibfnamefont {A.}~\bibnamefont {Omran}}, \bibinfo {author}
  {\bibfnamefont {T.~T.}\ \bibnamefont {Wang}}, \bibinfo {author}
  {\bibfnamefont {S.}~\bibnamefont {Ebadi}}, \emph {et~al.},\ }\bibfield
  {title} {\bibinfo {title} {{Parallel Implementation of High-Fidelity
  Multiqubit Gates with Neutral Atoms}},\ }\href
  {https://doi.org/10.1103/PhysRevLett.123.170503} {\bibfield  {journal}
  {\bibinfo  {journal} {Physical Review Letters}\ }\textbf {\bibinfo {volume}
  {123}},\ \bibinfo {pages} {170503} (\bibinfo {year} {2019})}.\BibitemShut {Stop}%
\bibitem [{\citenamefont {Evered}\ \emph {et~al.}(2023)\citenamefont {Evered},
  \citenamefont {Bluvstein}, \citenamefont {Kalinowski}, \citenamefont {Ebadi},
  \citenamefont {Manovitz}, \citenamefont {Zhou} \emph
  {et~al.}}]{evered2023high}%
  \BibitemOpen
  \bibfield  {author} {\bibinfo {author} {\bibfnamefont {S.~J.}\ \bibnamefont
  {Evered}}, \bibinfo {author} {\bibfnamefont {D.}~\bibnamefont {Bluvstein}},
  \bibinfo {author} {\bibfnamefont {M.~W.}\ \bibnamefont {Kalinowski}},
  \bibinfo {author} {\bibfnamefont {S.}~\bibnamefont {Ebadi}}, \bibinfo
  {author} {\bibfnamefont {T.}~\bibnamefont {Manovitz}}, \bibinfo {author}
  {\bibfnamefont {H.}~\bibnamefont {Zhou}}, \emph {et~al.},\ }\bibfield
  {title} {\bibinfo {title} {{High-fidelity parallel entangling gates on a
  neutral-atom quantum computer}},\ }\href
  {https://doi.org/10.1038/s41586-023-06481-y} {\bibfield  {journal} {\bibinfo
  {journal} {Nature}\ }\textbf {\bibinfo {volume} {622}},\ \bibinfo {pages}
  {268} (\bibinfo {year} {2023})}.\BibitemShut {Stop}%
\bibitem [{\citenamefont {Jandura}\ and\ \citenamefont
  {Pupillo}(2022)}]{jandura2022time}%
  \BibitemOpen
  \bibfield  {author} {\bibinfo {author} {\bibfnamefont {S.}~\bibnamefont
  {Jandura}}\ and\ \bibinfo {author} {\bibfnamefont {G.}~\bibnamefont
  {Pupillo}},\ }\bibfield  {title} {\bibinfo {title} {{Time-Optimal Two- and
  Three-Qubit Gates for Rydberg Atoms}},\ }\href
  {https://doi.org/10.22331/q-2022-05-13-712} {\bibfield  {journal} {\bibinfo
  {journal} {Quantum}\ }\textbf {\bibinfo {volume} {6}},\ \bibinfo {pages}
  {712} (\bibinfo {year} {2022})}.\BibitemShut {Stop}%
\bibitem [{\citenamefont {Deist}\ \emph {et~al.}(2022)\citenamefont {Deist},
  \citenamefont {Lu}, \citenamefont {Ho}, \citenamefont {Pasha}, \citenamefont
  {Zeiher}, \citenamefont {Yan},\ and\ \citenamefont
  {Stamper-Kurn}}]{deist2022mid}%
  \BibitemOpen
  \bibfield  {author} {\bibinfo {author} {\bibfnamefont {E.}~\bibnamefont
  {Deist}}, \bibinfo {author} {\bibfnamefont {Y.-H.}\ \bibnamefont {Lu}},
  \bibinfo {author} {\bibfnamefont {J.}~\bibnamefont {Ho}}, \bibinfo {author}
  {\bibfnamefont {M.~K.}\ \bibnamefont {Pasha}}, \bibinfo {author}
  {\bibfnamefont {J.}~\bibnamefont {Zeiher}}, \bibinfo {author} {\bibfnamefont
  {Z.}~\bibnamefont {Yan}},\ and\ \bibinfo {author} {\bibfnamefont
  {D.}~\bibnamefont {Stamper-Kurn}},\ }\bibfield  {title} {\bibinfo {title}
  {{Mid-Circuit Cavity Measurement in a Neutral Atom Array}},\ }\href
  {https://doi.org/10.1103/PhysRevLett.129.203602} {\bibfield  {journal}
  {\bibinfo  {journal} {Physical Review Letters}\ }\textbf {\bibinfo {volume}
  {129}},\ \bibinfo {pages} {203602} (\bibinfo {year} {2022})}.\BibitemShut {Stop}%
\bibitem [{\citenamefont {Lis}\ \emph {et~al.}(2023)\citenamefont {Lis},
  \citenamefont {Senoo}, \citenamefont {McGrew}, \citenamefont {Ronchen},
  \citenamefont {Jenkins},\ and\ \citenamefont {Kaufman}}]{lis2023midcircuit}%
  \BibitemOpen
  \bibfield  {author} {\bibinfo {author} {\bibfnamefont {J.~W.}\ \bibnamefont
  {Lis}}, \bibinfo {author} {\bibfnamefont {A.}~\bibnamefont {Senoo}}, \bibinfo
  {author} {\bibfnamefont {W.~F.}\ \bibnamefont {McGrew}}, \bibinfo {author}
  {\bibfnamefont {F.}~\bibnamefont {Ronchen}}, \bibinfo {author} {\bibfnamefont
  {A.}~\bibnamefont {Jenkins}},\ and\ \bibinfo {author} {\bibfnamefont {A.~M.}\
  \bibnamefont {Kaufman}},\ }\bibfield  {title} {\bibinfo {title} {{Midcircuit
  Operations Using the omg Architecture in Neutral Atom Arrays}},\ }\href
  {https://doi.org/10.1103/PhysRevX.13.041035} {\bibfield  {journal} {\bibinfo
  {journal} {Physical Review X}\ }\textbf {\bibinfo {volume} {13}},\ \bibinfo
  {pages} {041035} (\bibinfo {year} {2023})}.\BibitemShut {Stop}%
\bibitem [{\citenamefont {Norcia}\ \emph {et~al.}(2023)\citenamefont {Norcia},
  \citenamefont {Cairncross}, \citenamefont {Barnes}, \citenamefont
  {Battaglino}, \citenamefont {Brown}, \citenamefont {Brown} \emph
  {et~al.}}]{norcia2023midcircuit}%
  \BibitemOpen
  \bibfield  {author} {\bibinfo {author} {\bibfnamefont {M.~A.}\ \bibnamefont
  {Norcia}}, \bibinfo {author} {\bibfnamefont {W.~B.}\ \bibnamefont
  {Cairncross}}, \bibinfo {author} {\bibfnamefont {K.}~\bibnamefont {Barnes}},
  \bibinfo {author} {\bibfnamefont {P.}~\bibnamefont {Battaglino}}, \bibinfo
  {author} {\bibfnamefont {A.~L.}\ \bibnamefont {Brown}}, \bibinfo {author}
  {\bibfnamefont {M.~O.}\ \bibnamefont {Brown}}, \emph {et~al.},\ }\bibfield
  {title} {\bibinfo {title} {{Midcircuit Qubit Measurement and Rearrangement in
  a $^{171}$Yb Atomic Array}},\ }\href
  {https://doi.org/10.1103/PhysRevX.13.041034} {\bibfield  {journal} {\bibinfo
  {journal} {Physical Review X}\ }\textbf {\bibinfo {volume} {13}},\ \bibinfo
  {pages} {041034} (\bibinfo {year} {2023})}.\BibitemShut {Stop}%
\bibitem [{\citenamefont {Bluvstein}\ \emph {et~al.}(2026)\citenamefont
  {Bluvstein}, \citenamefont {Geim}, \citenamefont {Li}, \citenamefont
  {Evered}, \citenamefont {Ataides}, \citenamefont {Baranes} \emph
  {et~al.}}]{bluvstein2025fault}%
  \BibitemOpen
  \bibfield  {author} {\bibinfo {author} {\bibfnamefont {D.}~\bibnamefont
  {Bluvstein}}, \bibinfo {author} {\bibfnamefont {A.~A.}\ \bibnamefont {Geim}},
  \bibinfo {author} {\bibfnamefont {S.~H.}\ \bibnamefont {Li}}, \bibinfo
  {author} {\bibfnamefont {S.~J.}\ \bibnamefont {Evered}}, \bibinfo {author}
  {\bibfnamefont {J.~P.~B.}\ \bibnamefont {Ataides}}, \bibinfo {author}
  {\bibfnamefont {G.}~\bibnamefont {Baranes}}, \emph {et~al.},\ }\bibfield
  {title} {\bibinfo {title} {{A fault-tolerant neutral-atom architecture for
  universal quantum computation}},\ }\href
  {https://doi.org/10.1038/s41586-025-09848-5} {\bibfield  {journal} {\bibinfo
  {journal} {Nature}\ }\textbf {\bibinfo {volume} {649}},\ \bibinfo {pages}
  {39} (\bibinfo {year} {2026})}.\BibitemShut {Stop}%
\bibitem [{\citenamefont {Graham}\ \emph {et~al.}(2023)\citenamefont {Graham},
  \citenamefont {Phuttitarn}, \citenamefont {Chinnarasu}, \citenamefont {Song},
  \citenamefont {Poole}, \citenamefont {Jooya}, \citenamefont {Scott},
  \citenamefont {Scott}, \citenamefont {Eichler},\ and\ \citenamefont
  {Saffman}}]{graham2023midcircuit}%
  \BibitemOpen
  \bibfield  {author} {\bibinfo {author} {\bibfnamefont {T.~M.}\ \bibnamefont
  {Graham}}, \bibinfo {author} {\bibfnamefont {L.}~\bibnamefont {Phuttitarn}},
  \bibinfo {author} {\bibfnamefont {R.}~\bibnamefont {Chinnarasu}}, \bibinfo
  {author} {\bibfnamefont {Y.}~\bibnamefont {Song}}, \bibinfo {author}
  {\bibfnamefont {C.}~\bibnamefont {Poole}}, \bibinfo {author} {\bibfnamefont
  {K.}~\bibnamefont {Jooya}}, \bibinfo {author} {\bibfnamefont {J.~J.}\
  \bibnamefont {Scott}}, \bibinfo {author} {\bibfnamefont {A.}~\bibnamefont
  {Scott}}, \bibinfo {author} {\bibfnamefont {P.}~\bibnamefont {Eichler}},\
  and\ \bibinfo {author} {\bibfnamefont {M.~E.}\ \bibnamefont {Saffman}},\
  }\bibfield  {title} {\bibinfo {title} {{Midcircuit Measurements on a
  Single-Species Neutral Alkali Atom Quantum Processor}},\ }\href
  {https://doi.org/10.1103/PhysRevX.13.041051} {\bibfield  {journal} {\bibinfo
  {journal} {Physical Review X}\ }\textbf {\bibinfo {volume} {13}},\ \bibinfo
  {pages} {041051} (\bibinfo {year} {2023})}.\BibitemShut {Stop}%
\bibitem [{\citenamefont {Singh}\ \emph {et~al.}(2023)\citenamefont {Singh},
  \citenamefont {Bradley}, \citenamefont {Anand}, \citenamefont {Ramesh},
  \citenamefont {White},\ and\ \citenamefont {Bernien}}]{singh2023mid}%
  \BibitemOpen
  \bibfield  {author} {\bibinfo {author} {\bibfnamefont {K.}~\bibnamefont
  {Singh}}, \bibinfo {author} {\bibfnamefont {C.~E.}\ \bibnamefont {Bradley}},
  \bibinfo {author} {\bibfnamefont {S.}~\bibnamefont {Anand}}, \bibinfo
  {author} {\bibfnamefont {V.}~\bibnamefont {Ramesh}}, \bibinfo {author}
  {\bibfnamefont {R.}~\bibnamefont {White}},\ and\ \bibinfo {author}
  {\bibfnamefont {H.}~\bibnamefont {Bernien}},\ }\bibfield  {title} {\bibinfo
  {title} {{Mid-circuit correction of correlated phase errors using an array of
  spectator qubits}},\ }\href {https://doi.org/10.1126/science.ade5337}
  {\bibfield  {journal} {\bibinfo  {journal} {Science}\ }\textbf {\bibinfo
  {volume} {380}},\ \bibinfo {pages} {1265} (\bibinfo {year}
  {2023})}.\BibitemShut {Stop}%
\bibitem [{\citenamefont {Laflamme}\ \emph {et~al.}(1996)\citenamefont
  {Laflamme}, \citenamefont {Miquel}, \citenamefont {Paz},\ and\ \citenamefont
  {Zurek}}]{laflamme1996perfect}%
  \BibitemOpen
  \bibfield  {author} {\bibinfo {author} {\bibfnamefont {R.}~\bibnamefont
  {Laflamme}}, \bibinfo {author} {\bibfnamefont {C.}~\bibnamefont {Miquel}},
  \bibinfo {author} {\bibfnamefont {J.~P.}\ \bibnamefont {Paz}},\ and\ \bibinfo
  {author} {\bibfnamefont {W.~H.}\ \bibnamefont {Zurek}},\ }\bibfield  {title}
  {\bibinfo {title} {{Perfect Quantum Error Correcting Code}},\ }\href
  {https://doi.org/10.1103/PhysRevLett.77.198} {\bibfield  {journal} {\bibinfo
  {journal} {Physical Review Letters}\ }\textbf {\bibinfo {volume} {77}},\
  \bibinfo {pages} {198} (\bibinfo {year} {1996})}.\BibitemShut {Stop}%
\bibitem [{\citenamefont {Cain}\ \emph {et~al.}(2024)\citenamefont {Cain},
  \citenamefont {Zhao}, \citenamefont {Zhou}, \citenamefont {Meister},
  \citenamefont {Ataides}, \citenamefont {Jaffe}, \citenamefont {Bluvstein},\
  and\ \citenamefont {Lukin}}]{cain2024correlated}%
  \BibitemOpen
  \bibfield  {author} {\bibinfo {author} {\bibfnamefont {M.}~\bibnamefont
  {Cain}}, \bibinfo {author} {\bibfnamefont {C.}~\bibnamefont {Zhao}}, \bibinfo
  {author} {\bibfnamefont {H.}~\bibnamefont {Zhou}}, \bibinfo {author}
  {\bibfnamefont {N.}~\bibnamefont {Meister}}, \bibinfo {author} {\bibfnamefont
  {J.~P.~B.}\ \bibnamefont {Ataides}}, \bibinfo {author} {\bibfnamefont
  {A.~M.}\ \bibnamefont {Jaffe}}, \bibinfo {author} {\bibfnamefont
  {D.}~\bibnamefont {Bluvstein}},\ and\ \bibinfo {author} {\bibfnamefont
  {M.~D.}\ \bibnamefont {Lukin}},\ }\bibfield  {title} {\bibinfo {title}
  {{Correlated Decoding of Logical Algorithms with Transversal Gates}},\ }\href
  {https://doi.org/10.1103/PhysRevLett.133.240602} {\bibfield  {journal}
  {\bibinfo  {journal} {Physical Review Letters}\ }\textbf {\bibinfo {volume}
  {133}},\ \bibinfo {pages} {240602} (\bibinfo {year} {2024})}.\BibitemShut {Stop}%
\bibitem [{\citenamefont {Gidney}(2021)}]{gidney2021stim}%
  \BibitemOpen
  \bibfield  {author} {\bibinfo {author} {\bibfnamefont {C.}~\bibnamefont
  {Gidney}},\ }\bibfield  {title} {\bibinfo {title} {{Stim: a fast stabilizer
  circuit simulator}},\ }\href {https://doi.org/10.22331/q-2021-07-06-497}
  {\bibfield  {journal} {\bibinfo  {journal} {Quantum}\ }\textbf {\bibinfo
  {volume} {5}},\ \bibinfo {pages} {497} (\bibinfo {year} {2021})}.\BibitemShut {Stop}%
\bibitem [{\citenamefont {Bravyi}\ \emph {et~al.}(2019)\citenamefont {Bravyi},
  \citenamefont {Browne}, \citenamefont {Calpin}, \citenamefont {Campbell},
  \citenamefont {Gosset},\ and\ \citenamefont {Howard}}]{bravyi2019simulation}%
  \BibitemOpen
  \bibfield  {author} {\bibinfo {author} {\bibfnamefont {S.}~\bibnamefont
  {Bravyi}}, \bibinfo {author} {\bibfnamefont {D.}~\bibnamefont {Browne}},
  \bibinfo {author} {\bibfnamefont {P.}~\bibnamefont {Calpin}}, \bibinfo
  {author} {\bibfnamefont {E.}~\bibnamefont {Campbell}}, \bibinfo {author}
  {\bibfnamefont {D.}~\bibnamefont {Gosset}},\ and\ \bibinfo {author}
  {\bibfnamefont {M.}~\bibnamefont {Howard}},\ }\bibfield  {title} {\bibinfo
  {title} {{Simulation of quantum circuits by low-rank stabilizer
  decompositions}},\ }\href {https://doi.org/10.22331/q-2019-09-02-181}
  {\bibfield  {journal} {\bibinfo  {journal} {Quantum}\ }\textbf {\bibinfo
  {volume} {3}},\ \bibinfo {pages} {181} (\bibinfo {year} {2019})}.\BibitemShut {Stop}%
\bibitem [{\citenamefont {Perlin}\ \emph {et~al.}(2020)\citenamefont {Perlin},
  \citenamefont {Qu},\ and\ \citenamefont {Rey}}]{perlin2020spin}%
  \BibitemOpen
  \bibfield  {author} {\bibinfo {author} {\bibfnamefont {M.~A.}\ \bibnamefont
  {Perlin}}, \bibinfo {author} {\bibfnamefont {C.}~\bibnamefont {Qu}},\ and\
  \bibinfo {author} {\bibfnamefont {A.~M.}\ \bibnamefont {Rey}},\ }\bibfield
  {title} {\bibinfo {title} {{Spin Squeezing with Short-Range Spin-Exchange
  Interactions}},\ }\href {https://doi.org/10.1103/PhysRevLett.125.223401}
  {\bibfield  {journal} {\bibinfo  {journal} {Physical Review Letters}\
  }\textbf {\bibinfo {volume} {125}},\ \bibinfo {pages} {223401} (\bibinfo
  {year} {2020})}.\BibitemShut {Stop}%
\bibitem [{\citenamefont {Block}\ \emph {et~al.}(2024)\citenamefont {Block},
  \citenamefont {Ye}, \citenamefont {Roberts}, \citenamefont {Chern},
  \citenamefont {Wu}, \citenamefont {Wang}, \citenamefont {Pollet},
  \citenamefont {Davis}, \citenamefont {Halperin},\ and\ \citenamefont
  {Yao}}]{block2024scalable}%
  \BibitemOpen
  \bibfield  {author} {\bibinfo {author} {\bibfnamefont {M.}~\bibnamefont
  {Block}}, \bibinfo {author} {\bibfnamefont {B.}~\bibnamefont {Ye}}, \bibinfo
  {author} {\bibfnamefont {B.}~\bibnamefont {Roberts}}, \bibinfo {author}
  {\bibfnamefont {S.}~\bibnamefont {Chern}}, \bibinfo {author} {\bibfnamefont
  {W.}~\bibnamefont {Wu}}, \bibinfo {author} {\bibfnamefont {Z.}~\bibnamefont
  {Wang}}, \bibinfo {author} {\bibfnamefont {L.}~\bibnamefont {Pollet}},
  \bibinfo {author} {\bibfnamefont {E.~J.}\ \bibnamefont {Davis}}, \bibinfo
  {author} {\bibfnamefont {B.~I.}\ \bibnamefont {Halperin}},\ and\ \bibinfo
  {author} {\bibfnamefont {N.~Y.}\ \bibnamefont {Yao}},\ }\bibfield  {title}
  {\bibinfo {title} {{Scalable spin squeezing from finite-temperature
  easy-plane magnetism}},\ }\href {https://doi.org/10.1038/s41567-024-02562-5}
  {\bibfield  {journal} {\bibinfo  {journal} {Nature Physics}\ }\textbf
  {\bibinfo {volume} {20}},\ \bibinfo {pages} {1575} (\bibinfo {year}
  {2024})}.\BibitemShut {Stop}%
\bibitem [{\citenamefont {Manetsch}\ \emph {et~al.}(2025)\citenamefont
  {Manetsch}, \citenamefont {Nomura}, \citenamefont {Bataille}, \citenamefont
  {Lv}, \citenamefont {Leung},\ and\ \citenamefont
  {Endres}}]{manetsch2025tweezer}%
  \BibitemOpen
  \bibfield  {author} {\bibinfo {author} {\bibfnamefont {H.~J.}\ \bibnamefont
  {Manetsch}}, \bibinfo {author} {\bibfnamefont {G.}~\bibnamefont {Nomura}},
  \bibinfo {author} {\bibfnamefont {E.}~\bibnamefont {Bataille}}, \bibinfo
  {author} {\bibfnamefont {X.}~\bibnamefont {Lv}}, \bibinfo {author}
  {\bibfnamefont {K.~H.}\ \bibnamefont {Leung}},\ and\ \bibinfo {author}
  {\bibfnamefont {M.}~\bibnamefont {Endres}},\ }\bibfield  {title} {\bibinfo
  {title} {{A tweezer array with 6,100 highly coherent atomic qubits}},\ }\href
  {https://doi.org/10.1038/s41586-025-09641-4} {\bibfield  {journal} {\bibinfo
  {journal} {Nature}\ }\textbf {\bibinfo {volume} {647}},\ \bibinfo {pages}
  {60} (\bibinfo {year} {2025})}.\BibitemShut {Stop}%
\bibitem [{\citenamefont {Wang}\ \emph {et~al.}(2026)\citenamefont {Wang},
  \citenamefont {Zhang}, \citenamefont {Zhang}, \citenamefont {Liao},
  \citenamefont {Wang}, \citenamefont {Tian}, \citenamefont {Ji}, \citenamefont
  {Wu}, \citenamefont {Ma}, \citenamefont {Qing}, \citenamefont {Li},
  \citenamefont {Zhang}, \citenamefont {Huang}, \citenamefont {Zhang},
  \citenamefont {Feng}, \citenamefont {Chen},\ and\ \citenamefont
  {Zhai}}]{wang2026trapping}%
  \BibitemOpen
  \bibfield  {author} {\bibinfo {author} {\bibfnamefont {Y.}~\bibnamefont
  {Wang}}, \bibinfo {author} {\bibfnamefont {Z.}~\bibnamefont {Zhang}},
  \bibinfo {author} {\bibfnamefont {T.}~\bibnamefont {Zhang}}, \bibinfo
  {author} {\bibfnamefont {Y.}~\bibnamefont {Liao}}, \bibinfo {author}
  {\bibfnamefont {H.}~\bibnamefont {Wang}}, \bibinfo {author} {\bibfnamefont
  {Y.}~\bibnamefont {Tian}}, \bibinfo {author} {\bibfnamefont {B.}~\bibnamefont
  {Ji}}, \bibinfo {author} {\bibfnamefont {Y.}~\bibnamefont {Wu}}, \bibinfo
  {author} {\bibfnamefont {L.}~\bibnamefont {Ma}}, \bibinfo {author}
  {\bibfnamefont {C.}~\bibnamefont {Qing}}, \bibinfo {author} {\bibfnamefont
  {C.}~\bibnamefont {Li}}, \bibinfo {author} {\bibfnamefont {W.}~\bibnamefont
  {Zhang}}, \bibinfo {author} {\bibfnamefont {Y.}~\bibnamefont {Huang}},
  \bibinfo {author} {\bibfnamefont {W.}~\bibnamefont {Zhang}}, \bibinfo
  {author} {\bibfnamefont {X.}~\bibnamefont {Feng}}, \bibinfo {author}
  {\bibfnamefont {W.}~\bibnamefont {Chen}},\ and\ \bibinfo {author}
  {\bibfnamefont {H.}~\bibnamefont {Zhai}},\ }\href
  {https://doi.org/10.48550/arXiv.2606.02715} {\bibinfo {title} {{Trapping
  11,000 Atoms in a Tweezer Array Generated by a Single Metasurface}}}
  (\bibinfo {year} {2026}),\ \Eprint {https://arxiv.org/abs/2606.02715}
  {arXiv:2606.02715} .\BibitemShut {Stop}%
\bibitem [{\citenamefont {Young}\ \emph {et~al.}(2023)\citenamefont {Young},
  \citenamefont {Muleady}, \citenamefont {Perlin}, \citenamefont {Kaufman},\
  and\ \citenamefont {Rey}}]{young2023enhancing}%
  \BibitemOpen
  \bibfield  {author} {\bibinfo {author} {\bibfnamefont {J.~T.}\ \bibnamefont
  {Young}}, \bibinfo {author} {\bibfnamefont {S.~R.}\ \bibnamefont {Muleady}},
  \bibinfo {author} {\bibfnamefont {M.~A.}\ \bibnamefont {Perlin}}, \bibinfo
  {author} {\bibfnamefont {A.~M.}\ \bibnamefont {Kaufman}},\ and\ \bibinfo
  {author} {\bibfnamefont {A.~M.}\ \bibnamefont {Rey}},\ }\bibfield  {title}
  {\bibinfo {title} {{Enhancing spin squeezing using soft-core interactions}},\
  }\href {https://doi.org/10.1103/PhysRevResearch.5.L012033} {\bibfield
  {journal} {\bibinfo  {journal} {Physical Review Research}\ }\textbf {\bibinfo
  {volume} {5}},\ \bibinfo {pages} {L012033} (\bibinfo {year}
  {2023})}.\BibitemShut {Stop}%
\bibitem [{\citenamefont {Steinert}\ \emph {et~al.}(2023)\citenamefont
  {Steinert}, \citenamefont {Osterholz}, \citenamefont {Eberhard},
  \citenamefont {Festa}, \citenamefont {Lorenz}, \citenamefont {Chen},
  \citenamefont {Trautmann},\ and\ \citenamefont
  {Gross}}]{steinert2023spatially}%
  \BibitemOpen
  \bibfield  {author} {\bibinfo {author} {\bibfnamefont {L.-M.}\ \bibnamefont
  {Steinert}}, \bibinfo {author} {\bibfnamefont {P.}~\bibnamefont {Osterholz}},
  \bibinfo {author} {\bibfnamefont {R.}~\bibnamefont {Eberhard}}, \bibinfo
  {author} {\bibfnamefont {L.}~\bibnamefont {Festa}}, \bibinfo {author}
  {\bibfnamefont {N.}~\bibnamefont {Lorenz}}, \bibinfo {author} {\bibfnamefont
  {Z.}~\bibnamefont {Chen}}, \bibinfo {author} {\bibfnamefont {A.}~\bibnamefont
  {Trautmann}},\ and\ \bibinfo {author} {\bibfnamefont {C.}~\bibnamefont
  {Gross}},\ }\bibfield  {title} {\bibinfo {title} {{Spatially Tunable Spin
  Interactions in Neutral Atom Arrays}},\ }\href
  {https://doi.org/10.1103/PhysRevLett.130.243001} {\bibfield  {journal}
  {\bibinfo  {journal} {Physical Review Letters}\ }\textbf {\bibinfo {volume}
  {130}},\ \bibinfo {pages} {243001} (\bibinfo {year} {2023})}.\BibitemShut {Stop}%
\bibitem [{\citenamefont {Bluvstein}\ \emph {et~al.}(2021)\citenamefont
  {Bluvstein}, \citenamefont {Omran}, \citenamefont {Levine}, \citenamefont
  {Keesling}, \citenamefont {Semeghini}, \citenamefont {Ebadi} \emph
  {et~al.}}]{bluvstein2021controlling}%
  \BibitemOpen
  \bibfield  {author} {\bibinfo {author} {\bibfnamefont {D.}~\bibnamefont
  {Bluvstein}}, \bibinfo {author} {\bibfnamefont {A.}~\bibnamefont {Omran}},
  \bibinfo {author} {\bibfnamefont {H.}~\bibnamefont {Levine}}, \bibinfo
  {author} {\bibfnamefont {A.}~\bibnamefont {Keesling}}, \bibinfo {author}
  {\bibfnamefont {G.}~\bibnamefont {Semeghini}}, \bibinfo {author}
  {\bibfnamefont {S.}~\bibnamefont {Ebadi}}, \emph {et~al.},\ }\bibfield
  {title} {\bibinfo {title} {{Controlling quantum many-body dynamics in driven
  Rydberg atom arrays}},\ }\href {https://doi.org/10.1126/science.abg2530}
  {\bibfield  {journal} {\bibinfo  {journal} {Science}\ }\textbf {\bibinfo
  {volume} {371}},\ \bibinfo {pages} {1355} (\bibinfo {year}
  {2021})}.\BibitemShut {Stop}%
\bibitem [{\citenamefont {Pan}\ and\ \citenamefont
  {Zhai}(2022)}]{pan2022composite}%
  \BibitemOpen
  \bibfield  {author} {\bibinfo {author} {\bibfnamefont {L.}~\bibnamefont
  {Pan}}\ and\ \bibinfo {author} {\bibfnamefont {H.}~\bibnamefont {Zhai}},\
  }\bibfield  {title} {\bibinfo {title} {{Composite spin approach to the
  blockade effect in Rydberg atom arrays}},\ }\href
  {https://doi.org/10.1103/PhysRevResearch.4.L032037} {\bibfield  {journal}
  {\bibinfo  {journal} {Physical Review Research}\ }\textbf {\bibinfo {volume}
  {4}},\ \bibinfo {pages} {L032037} (\bibinfo {year} {2022})}.\BibitemShut {Stop}%
\bibitem [{\citenamefont {Yao}\ \emph {et~al.}(2022)\citenamefont {Yao},
  \citenamefont {Pan}, \citenamefont {Liu},\ and\ \citenamefont
  {Zhai}}]{yao2022quantum}%
  \BibitemOpen
  \bibfield  {author} {\bibinfo {author} {\bibfnamefont {Z.}~\bibnamefont
  {Yao}}, \bibinfo {author} {\bibfnamefont {L.}~\bibnamefont {Pan}}, \bibinfo
  {author} {\bibfnamefont {S.}~\bibnamefont {Liu}},\ and\ \bibinfo {author}
  {\bibfnamefont {H.}~\bibnamefont {Zhai}},\ }\bibfield  {title} {\bibinfo
  {title} {{Quantum many-body scars and quantum criticality}},\ }\href
  {https://doi.org/10.1103/PhysRevB.105.125123} {\bibfield  {journal} {\bibinfo
   {journal} {Physical Review B}\ }\textbf {\bibinfo {volume} {105}},\ \bibinfo
  {pages} {125123} (\bibinfo {year} {2022})}.\BibitemShut {Stop}%
\bibitem [{\citenamefont {Cheng}\ \emph {et~al.}(2022)\citenamefont {Cheng},
  \citenamefont {Liu}, \citenamefont {Zheng}, \citenamefont {Zhang},\ and\
  \citenamefont {Zhai}}]{cheng2022tunable}%
  \BibitemOpen
  \bibfield  {author} {\bibinfo {author} {\bibfnamefont {Y.}~\bibnamefont
  {Cheng}}, \bibinfo {author} {\bibfnamefont {S.}~\bibnamefont {Liu}}, \bibinfo
  {author} {\bibfnamefont {W.}~\bibnamefont {Zheng}}, \bibinfo {author}
  {\bibfnamefont {P.}~\bibnamefont {Zhang}},\ and\ \bibinfo {author}
  {\bibfnamefont {H.}~\bibnamefont {Zhai}},\ }\bibfield  {title} {\bibinfo
  {title} {{Tunable Confinement-Deconfinement Transition in an Ultracold-Atom
  Quantum Simulator}},\ }\href {https://doi.org/10.1103/PRXQuantum.3.040317}
  {\bibfield  {journal} {\bibinfo  {journal} {PRX Quantum}\ }\textbf {\bibinfo
  {volume} {3}},\ \bibinfo {pages} {040317} (\bibinfo {year}
  {2022})}.\BibitemShut {Stop}%
\bibitem [{\citenamefont {Cheng}\ \emph {et~al.}(2023)\citenamefont {Cheng},
  \citenamefont {Li},\ and\ \citenamefont {Zhai}}]{cheng2023variational}%
  \BibitemOpen
  \bibfield  {author} {\bibinfo {author} {\bibfnamefont {Y.}~\bibnamefont
  {Cheng}}, \bibinfo {author} {\bibfnamefont {C.}~\bibnamefont {Li}},\ and\
  \bibinfo {author} {\bibfnamefont {H.}~\bibnamefont {Zhai}},\ }\bibfield
  {title} {\bibinfo {title} {{Variational approach to quantum spin liquid in a
  Rydberg atom simulator}},\ }\href {https://doi.org/10.1088/1367-2630/acc125}
  {\bibfield  {journal} {\bibinfo  {journal} {New Journal of Physics}\ }\textbf
  {\bibinfo {volume} {25}},\ \bibinfo {pages} {033010} (\bibinfo {year}
  {2023})}.\BibitemShut {Stop}%
\bibitem [{\citenamefont {Li}\ \emph {et~al.}(2024)\citenamefont {Li},
  \citenamefont {Liu}, \citenamefont {Wang}, \citenamefont {Zhang},
  \citenamefont {Li}, \citenamefont {Zhai},\ and\ \citenamefont
  {Gu}}]{li2024uncovering}%
  \BibitemOpen
  \bibfield  {author} {\bibinfo {author} {\bibfnamefont {C.}~\bibnamefont
  {Li}}, \bibinfo {author} {\bibfnamefont {S.}~\bibnamefont {Liu}}, \bibinfo
  {author} {\bibfnamefont {H.}~\bibnamefont {Wang}}, \bibinfo {author}
  {\bibfnamefont {W.}~\bibnamefont {Zhang}}, \bibinfo {author} {\bibfnamefont
  {Z.}~\bibnamefont {Li}}, \bibinfo {author} {\bibfnamefont {H.}~\bibnamefont
  {Zhai}},\ and\ \bibinfo {author} {\bibfnamefont {Y.}~\bibnamefont {Gu}},\
  }\bibfield  {title} {\bibinfo {title} {{Uncovering Emergent Spacetime
  Supersymmetry with Rydberg Atom Arrays}},\ }\href
  {https://doi.org/10.1103/PhysRevLett.133.223401} {\bibfield  {journal}
  {\bibinfo  {journal} {Physical Review Letters}\ }\textbf {\bibinfo {volume}
  {133}},\ \bibinfo {pages} {223401} (\bibinfo {year} {2024})}.\BibitemShut {Stop}%
\bibitem [{\citenamefont {Zhang}\ \emph {et~al.}(2025)\citenamefont {Zhang},
  \citenamefont {Wang}, \citenamefont {Zhang}, \citenamefont {Wang},
  \citenamefont {Du}, \citenamefont {Li}, \citenamefont {Wu}, \citenamefont
  {Li}, \citenamefont {Hu}, \citenamefont {Zhai},\ and\ \citenamefont
  {Chen}}]{zhang2025observation}%
  \BibitemOpen
  \bibfield  {author} {\bibinfo {author} {\bibfnamefont {T.}~\bibnamefont
  {Zhang}}, \bibinfo {author} {\bibfnamefont {H.}~\bibnamefont {Wang}},
  \bibinfo {author} {\bibfnamefont {W.}~\bibnamefont {Zhang}}, \bibinfo
  {author} {\bibfnamefont {Y.}~\bibnamefont {Wang}}, \bibinfo {author}
  {\bibfnamefont {A.}~\bibnamefont {Du}}, \bibinfo {author} {\bibfnamefont
  {Z.}~\bibnamefont {Li}}, \bibinfo {author} {\bibfnamefont {Y.}~\bibnamefont
  {Wu}}, \bibinfo {author} {\bibfnamefont {C.}~\bibnamefont {Li}}, \bibinfo
  {author} {\bibfnamefont {J.}~\bibnamefont {Hu}}, \bibinfo {author}
  {\bibfnamefont {H.}~\bibnamefont {Zhai}},\ and\ \bibinfo {author}
  {\bibfnamefont {W.}~\bibnamefont {Chen}},\ }\bibfield  {title} {\bibinfo
  {title} {{Observation of Near-Critical Kibble-Zurek Scaling in Rydberg Atom
  Arrays}},\ }\href {https://doi.org/10.1103/2gwz-65w1} {\bibfield  {journal}
  {\bibinfo  {journal} {Physical Review Letters}\ }\textbf {\bibinfo {volume}
  {135}},\ \bibinfo {pages} {093403} (\bibinfo {year} {2025})}.\BibitemShut {Stop}%
\bibitem [{\citenamefont {Ma}\ \emph {et~al.}(2023)\citenamefont {Ma},
  \citenamefont {Liu}, \citenamefont {Peng}, \citenamefont {Zhang},
  \citenamefont {Jandura}, \citenamefont {Claes}, \citenamefont {Burgers},
  \citenamefont {Pupillo}, \citenamefont {Puri},\ and\ \citenamefont
  {Thompson}}]{ma2023high}%
  \BibitemOpen
  \bibfield  {author} {\bibinfo {author} {\bibfnamefont {S.}~\bibnamefont
  {Ma}}, \bibinfo {author} {\bibfnamefont {G.}~\bibnamefont {Liu}}, \bibinfo
  {author} {\bibfnamefont {P.}~\bibnamefont {Peng}}, \bibinfo {author}
  {\bibfnamefont {B.}~\bibnamefont {Zhang}}, \bibinfo {author} {\bibfnamefont
  {S.}~\bibnamefont {Jandura}}, \bibinfo {author} {\bibfnamefont
  {J.}~\bibnamefont {Claes}}, \bibinfo {author} {\bibfnamefont {A.~P.}\
  \bibnamefont {Burgers}}, \bibinfo {author} {\bibfnamefont {G.}~\bibnamefont
  {Pupillo}}, \bibinfo {author} {\bibfnamefont {S.}~\bibnamefont {Puri}},\ and\
  \bibinfo {author} {\bibfnamefont {J.~D.}\ \bibnamefont {Thompson}},\
  }\bibfield  {title} {\bibinfo {title} {{High-fidelity gates and mid-circuit
  erasure conversion in an atomic qubit}},\ }\href
  {https://doi.org/10.1038/s41586-023-06438-1} {\bibfield  {journal} {\bibinfo
  {journal} {Nature}\ }\textbf {\bibinfo {volume} {622}},\ \bibinfo {pages}
  {279} (\bibinfo {year} {2023})}.\BibitemShut {Stop}%
\bibitem [{\citenamefont {Scholl}\ \emph {et~al.}(2023)\citenamefont {Scholl},
  \citenamefont {Shaw}, \citenamefont {Tsai}, \citenamefont {Finkelstein},
  \citenamefont {Choi},\ and\ \citenamefont {Endres}}]{scholl2023erasure}%
  \BibitemOpen
  \bibfield  {author} {\bibinfo {author} {\bibfnamefont {P.}~\bibnamefont
  {Scholl}}, \bibinfo {author} {\bibfnamefont {A.~L.}\ \bibnamefont {Shaw}},
  \bibinfo {author} {\bibfnamefont {R.~B.-S.}\ \bibnamefont {Tsai}}, \bibinfo
  {author} {\bibfnamefont {R.}~\bibnamefont {Finkelstein}}, \bibinfo {author}
  {\bibfnamefont {J.}~\bibnamefont {Choi}},\ and\ \bibinfo {author}
  {\bibfnamefont {M.}~\bibnamefont {Endres}},\ }\bibfield  {title} {\bibinfo
  {title} {{Erasure conversion in a high-fidelity Rydberg quantum simulator}},\
  }\href {https://doi.org/10.1038/s41586-023-06516-4} {\bibfield  {journal}
  {\bibinfo  {journal} {Nature}\ }\textbf {\bibinfo {volume} {622}},\ \bibinfo
  {pages} {273} (\bibinfo {year} {2023})}.\BibitemShut {Stop}%
\bibitem [{\citenamefont {Chao}\ and\ \citenamefont
  {Reichardt}(2018{\natexlab{a}})}]{chao2018fault}%
  \BibitemOpen
  \bibfield  {author} {\bibinfo {author} {\bibfnamefont {R.}~\bibnamefont
  {Chao}}\ and\ \bibinfo {author} {\bibfnamefont {B.~W.}\ \bibnamefont
  {Reichardt}},\ }\bibfield  {title} {\bibinfo {title} {{Fault-tolerant quantum
  computation with few qubits}},\ }\href
  {https://doi.org/10.1038/s41534-018-0085-z} {\bibfield  {journal} {\bibinfo
  {journal} {npj Quantum Information}\ }\textbf {\bibinfo {volume} {4}},\
  \bibinfo {pages} {42} (\bibinfo {year} {2018}{\natexlab{a}})}.\BibitemShut {Stop}%
\bibitem [{\citenamefont {Chao}\ and\ \citenamefont
  {Reichardt}(2018{\natexlab{b}})}]{chao2018quantum}%
  \BibitemOpen
  \bibfield  {author} {\bibinfo {author} {\bibfnamefont {R.}~\bibnamefont
  {Chao}}\ and\ \bibinfo {author} {\bibfnamefont {B.~W.}\ \bibnamefont
  {Reichardt}},\ }\bibfield  {title} {\bibinfo {title} {{Quantum Error
  Correction with Only Two Extra Qubits}},\ }\href
  {https://doi.org/10.1103/PhysRevLett.121.050502} {\bibfield  {journal}
  {\bibinfo  {journal} {Physical Review Letters}\ }\textbf {\bibinfo {volume}
  {121}},\ \bibinfo {pages} {050502} (\bibinfo {year}
  {2018}{\natexlab{b}})}.\BibitemShut {Stop}%
\bibitem[Evered et al.(2026)]{evered2026nonlocal}
\BibitemOpen S. J. Evered, M. Xu, S. H. Li, A. A. Geim, J. P. Bonilla Ataides, M. Kalinowski, D. Bluvstein, N. Maskara, C. Kokail, M. Greiner, V. Vuleti\'c, and M. D. Lukin, High-fidelity entangling gates and nonlocal circuits with neutral atoms, \href{https://doi.org/10.1126/science.aei1312}{Science \textbf{394}, 197 (2026)}.\BibitemShut{Stop}%
\bibitem[Chiu et al.(2025)]{chiu2025continuous}
\BibitemOpen N.-C. Chiu, E. C. Trapp, J. Guo, M. H. Abobeih, L. M. Stewart, S. Hollerith, et al., Continuous operation of a coherent 3,000-qubit system, \href{https://doi.org/10.1038/s41586-025-09596-6}{Nature \textbf{646}, 1075 (2025)}.\BibitemShut{Stop}%
\bibitem[Hoeffding(1963)]{hoeffding1963probability}
\BibitemOpen W. Hoeffding, Probability inequalities for sums of bounded random variables, \href{https://doi.org/10.1080/01621459.1963.10500830}{J. Am. Stat. Assoc. \textbf{58}, 13 (1963)}.\BibitemShut{Stop}%

\bibitem[Daguerre et al.(2025)]{daguerre2025experimental}
\BibitemOpen L. Daguerre, R. Blume-Kohout, N. C. Brown, D. Hayes, and I. H. Kim, Experimental demonstration of high-fidelity logical magic states from code switching, \href{https://doi.org/10.1103/dck4-x9c2}{Phys. Rev. X \textbf{15}, 041008 (2025)}.\BibitemShut{Stop}%

\end{thebibliography}
\end{document}